\documentclass[12pt,a4paper]{article}
\usepackage{cmap}
\usepackage[T2A]{fontenc}
\usepackage[cp1251]{inputenc}
\usepackage[english]{babel}
\usepackage{amsfonts,amsmath,amssymb,amsthm,longtable,hhline}
\usepackage{bm}
\usepackage{cite}
\usepackage{soul}
\usepackage{color,xcolor}
\usepackage{hyperref}
\usepackage{multicol}
\usepackage{tabularx,caption,multirow}
\makeatletter
\renewcommand{\@biblabel}[1]{#1.} 
\makeatother
\theoremstyle{remark}

\newcounter{urav}[section]
\newcounter{resh}[urav]

\let\ds=\displaystyle
\let\ts=\textstyle
\let\bl=\bigl \let\br=\bigr
\let\Bl=\Bigl \let\Br=\Bigr
\let\BL=\biggl \let\BR=\biggr
\def\arb{is an arbitrary constant}
\def\arbs{are arbitrary constants}
\def\arbf{is an arbitrary function}
\def\arbfs{are arbitrary functions}
\def\fracskip{\mskip 1mu \relax}
\def\nfrac#1#2{{\fracskip#1\fracskip\over\fracskip#2\fracskip}}
\def\dfrac#1#2{{\ds\nfrac{#1}{#2}}}
\def\tfrac#1#2{{\ts\nfrac{#1}{#2}}}
\let\frac=\nfrac
\def\pd#1#2{\dfrac{\partial#1}{\partial#2}}
\def\pdd#1#2#3{\ifx#2#3\pd{^2#1}{#2^2}\else\pd{^2#1}{#2\partial#3}\fi }

\newcommand{\clhp}[1]{\colorbox{yellow}{\parbox{\textwidth}{#1}}}%

\begin{document}
\large 

\section*{}

\centerline{\Large\bf Functional separation of variables in nonlinear PDEs:}
\centerline{\Large\bf General approach, new solutions of diffusion-type equations}

\bigskip

\centerline{Andrei D. Polyanin}
\bigskip
\centerline{\it Ishlinsky Institute for Problems in Mechanics, Russian Academy of Sciences,}
\centerline{\it 101 Vernadsky Avenue, bldg~1, 119526 Moscow, Russia}
\bigskip
\bigskip

\textbf{Abstract.} The study gives a brief overview of existing modifications of the method of functional separation of variables for nonlinear PDEs. It proposes a more general approach to the
construction of exact solutions to nonlinear equations of applied mathematics and
mathematical physics, based on a special transformation with an integral term and the
generalized splitting principle. The effectiveness of this approach is illustrated by nonlinear diffusion-type equations that contain reaction and
convective terms with variable coefficients. The focus is on equations of a fairly general form that depend on one, two or three arbitrary functions (such nonlinear
PDEs are most difficult to analyze and find exact solutions).\let\thefootnote\relax\footnotetext{
\hskip-20pt\clhp{$^*$ This is a preprint of an article that will be published in the journal
\textit{Mathematics}, 2020, Vol. 8, No. 1, 90; doi:10.3390/math8010090}}
A number of new functional separable solutions and
generalized traveling wave solutions are described (more than 30 exact solutions have been presented in total). It is shown that the method of functional separation
of variables can, in certain cases, be more effective than (i) the
nonclassical method of symmetry reductions based on an invariant surface condition, and (ii) the method of differential constraints based on a single
differential constraint. The exact solutions obtained can be used to test various numerical and approximate analytical methods of mathematical physics and mechanics.
\bigskip

\textsl{Keywords:\/}
functional separation of variables;
generalized separation of variables;
exact solutions;
nonlinear reaction-diffusion equations;
nonlinear partial differential equations;
splitting principle;
nonclassical method of symmetry reductions;
invariant surface condition;
differential constraints

\tableofcontents

\section{Introduction}\label{s:1}

\subsection{A brief overview of modifications of the method of functional\\ separation of variables}\label{ss:1.1}

The methods of generalized and functional separation of variables (and their various modifications) are
among the most effective methods for constructing exact solutions to various classes of nonlinear equations of mathematical physics and mechanics
(including partial differential equations of fairly general forms that involve arbitrary functions).
In \cite{mil1989,grun1992,mil1993,zhd1994,gal1995,gal1995b,and1998,doy1998,puc2000,qu2000,pol2001,est2002,est2002a,zha2003,zha2004a,zha2004b,zha2005,pol2005,gal2007,hu2007,pen2007,
jia2008,pol2012,bar2013,pol2014,pol2015,pol2016,pol2016a,pol2016b,pol2019a,pol2019b,zhur2000,pol2019c,pol2019d,pol2019e,pol2019f}, many exact solutions to equations of heat and mass transfer theory, wave theory,
hydrodynamics, gas dynamics, nonlinear optics, and mathematical biology were obtained using these methods.

To be specific, we will consider nonlinear PDEs of mathematical physics with two independent variables
\begin{align}
F(x,t,u_x,u_t,u_{xx},u_{xt},u_{tt},\dots)=0, \label{eq:01}
\end{align}
where $u=u(x,t)$ is the unknown function.

The methods of generalized and functional separation of variables are based on setting a priori a
structural form of $u$ that depends on several free functions
(the specific form of these functions is determined subsequently by analyzing the arising functional differential equations).

Exact solutions in the form of the sum or product of two functions that depend on different arguments,
$$
u = \xi(x)+\eta(t)\quad \ \text{or} \quad \ u = \xi(x)\eta(t),
$$
are called ordinary separable solutions.
Examples of nonlinear PDEs with such solutions can be found in \cite{qu2000,pol2005,pol2012}.

Often (in a narrow sense) the term `solution with functional separation of variables'
(or `functional separable solution') is used for exact solutions of the form (e.g., see
\cite{mil1989,grun1992,mil1993,zhd1994,and1998,doy1998,pol2005,jia2008,pol2012})
\begin{align}
u=\varphi(z),\quad \ z=\xi(x)+\eta(t),
\label{eq:a00}
\end{align}
where the functions $\varphi(z)$, $\xi(x)$, and $\eta(t)$ are determined in a subsequent analysis.
Sometimes the external function $\varphi(z)$ is specified from a priori considerations, while the internal functions $\xi(x)$ and $\eta(t)$ are to be found
\cite{gal2007,pol2012}.

Importantly, in functional separation of variables, the search for exact solutions of the form
$u=\varphi(\xi(x)\eta(t))$ and $u=\varphi(\xi(x)+\eta(t))$ leads to the same results, since
$\varphi(\xi(x)\eta(t))=\varphi_1(\xi_1(x)+\eta_1(t))$, where
$\varphi_1(z)=\varphi(e^z)$, $\varphi_1(x)=\ln\varphi(x)$, and $\psi_1(t)=\ln\psi(t)$
(here, without loss of generality, it is assumed that $\varphi>0$ and $\psi>0$).

Generalized traveling wave solutions of the form $u=\varphi(z)$, where $z=\xi(t)x+\eta(t)$
are treated as solutions with functional separation of variables \cite{pol2005,pol2012}.

In \cite{qu2000,zha2005,hu2007,jia2008},
the representation of functional separable solutions in implicit form
\begin{align}
\psi(u)=\xi(x)+\eta(t),
\label{eq:b00}
\end{align}
was used.
All three functions $\psi(u)$, $\xi(x)$, and $\eta(t)$ were to be found.
More complex functionally separable solutions of the form
$u=\varphi(z)$, where $z=\eta_1(t)\xi(x)+\eta_2(t)$,
were considered in \cite{puc2000,gal1989,pol2012}.

The studies \cite{pol2019c,pol2019d,pol2019e} described a new direct method for constructing exact solutions with functional separation of variables.
It is based on an implicit integral representation of solutions in the form
\begin{align}
\int \zeta(u)\,du=\xi_1(x)\eta(t)+\xi_2(x),
\label{eq:00}
\end{align}
where the functions $\zeta(u)$, $\xi_1(x)$, $\xi_2(x)$, and $\eta(t)$ are determined by the splitting method in the subsequent analysis.
This method allowed to find more than 40 exact solutions of nonlinear reaction-diffusion equations and wave type equations with variable
coefficients involving one or more arbitrary functions.
In \cite{pol2019f}, it was shown that some of the solutions given in \cite{pol2019d,pol2019e} cannot be obtained using
the nonclassical method of symmetry reductions
\cite{blu1969,lev1989,nuc1992,pic1992a,cla1995,olv1996,cla1997,kap2003} (see also \cite{pol2005,pol2012})
based on the use of the invariant surface condition (a first-order differential constraint equivalent to the relation \eqref{eq:00}).

Note that constructing solutions in implicit form with the integral term \eqref{eq:00}
often allows us to reduce the order of the resulting functional differential equations \cite{pol2019c,pol2019d}.

In the general case, the term `functional separable solution' with regard to nonlinear PDEs \eqref{eq:01} will be used for exact solutions that can be represented as
\begin{align}
u=\varphi(z),\quad \ z=Q(x,t),
\label{eq:000}
\end{align}
where the desired functions $\varphi(z)$ and $Q(x,t)$ are described respectively by overdetermined systems of ODEs and PDEs.
In the simplest cases, each of these functions can be described by a single equation.
Representation \eqref{eq:000} was used in \cite{pol2019a,pol2019b,zhur2000} to construct exact solutions with functional separation of variables
to some classes of nonlinear reaction-diffusion, convective-diffusion, and wave type equations.

It is necessary to distinguish between direct and indirect functional separation of variables based on one of the representations of solutions
\eqref{eq:a00}, \eqref{eq:b00}, \eqref{eq:00}, or \eqref{eq:000}.
At the first stage of direct functional separation of variables, the representation of solution is substituted into the original PDE,
after which the resulting equation is analyzed (e.g., see \cite{pol2012,pol2019a,pol2019b,pol2019c,pol2019d,pol2019e}).
At the first stage of indirect functional separation of variables, the representation of solution is replaced by one or more equivalent
differential constraints, and then the overdetermined system of PDEs obtained in this way is analyzed for compatibility (e.g., see \cite{puc2000,hu2007,jia2008,pol2019f}).

To construct exact solutions of nonlinear partial differential equations, this paper
proposes to use a direct method based on
a special transformation with an integral term as well as the generalized splitting principle.
This approach is technically simpler and more convenient than finding a solution in the form \eqref{eq:000};
it generalizes the dependence \eqref{eq:00} and allows one to find various solutions
in a uniform manner without specifying their structure a priori.

\subsection{The concept of `exact solution' for nonlinear PDEs}\label{ss:1.2}

In what follows, the term `exact solution' with regard to nonlinear
partial differential equations is used in the following cases:

\begin{itemize}

\item[(i)] the solution is expressible in terms of elementary functions;

\item[(ii)] the solution is expressible in terms of elementary functions,
functions included in the equation in question, and indefinite or/and definite integrals;

\item[(iii)] the solution is expressible in terms of solutions to ordinary
differential equations or systems of such equations.

Combinations of cases (i) and (iii) as well as (ii) and (iii) are also allowed.

\end{itemize}

Case (i) is specially isolated from the more general case (ii) as its simplest variant.
In cases (i) and (ii), an exact solution can be represented in explicit, implicit or parametric form.

\section{Direct functional separation of variables. General\\ approach}\label{s:2}

\subsection{Method description. The generalized splitting principle}\label{ss:2.1}

To construct exact solutions of equation \eqref{eq:01}, we first introduce a new dependent variable $\vartheta$ using the nonlinear transformation
\begin{align}
\vartheta=\int \zeta(u)\,du. \label{eq:02}
\end{align}
Both functions $\vartheta=\vartheta(x,t)$ and $\zeta=\zeta(u)$
will be found simultaneously in the subsequent analysis.
Once these functions are determined, the integral relation \eqref{eq:02}
will specify an exact solution of the equation in question in implicit form
(in some cases, the solution may be represented explicitly).

Differentiating \eqref{eq:02} with respect to the independent variables, we find the partial derivatives
\begin{align}
u_x=\frac{\vartheta_x}\zeta,\quad u_t=\frac{\vartheta_t}\zeta,\quad
u_{xx}=\frac{\vartheta_{xx}}\zeta-\frac{\vartheta_x^2\zeta'_u}{\zeta^3},\quad
u_{xt}=\frac{\vartheta_{xt}}\zeta-\frac{\vartheta_x\vartheta_t\zeta'_u}{\zeta^3},\quad \dots
\label{eq:03}
\end{align}
We assume that after substituting expressions \eqref{eq:03} into \eqref{eq:01},
the resulting equation can be converted to the following form:
\begin{equation}
\sum^N_{n=1}\Phi_n\Psi_n=0,
\label{eq:04}
\end{equation}
where
\begin{equation}
\begin{aligned}
\Phi_n&=\Phi_n(x,t,\vartheta_x,\vartheta_t,\vartheta_{xx},\dots),\\
\Psi_n&=\Psi_n(u,\zeta,\zeta'_u,\zeta''_{uu},\dots).
\end{aligned}
\label{eq:05}
\end{equation}

To construct exact solutions of equation \eqref{eq:04}--\eqref{eq:05}, we use the splitting principle described below.

\medbreak
\textit{The generalized splitting principle}.
We consider linear combinations of two sets of elements $\{\Phi_j\}$ and $\{\Psi_j\}$
included in \eqref{eq:04}, which are connected by relations
\begin{equation}
\begin{aligned}
&\sum^{N}_{n=1}\alpha_{ni}\Phi_n=0,\quad \ i=1,\dots,l;\\
&\sum^{N}_{n=1}\beta_{nj}\Psi_n=0,\quad \ j=1,\dots,m,
\end{aligned}
\label{eq:06}
\end{equation}
where $1\le l\le N-1$ and $1\le m\le N-1$.
The constants $\alpha_{ni}$ and $\beta_{nj}$ in \eqref{eq:06} are chosen so that the bilinear equality \eqref{eq:04}
is satisfied identically (this can always be done as shown below).
Importantly, relations \eqref{eq:06} are purely algebraic
in nature and are independent of any particular expressions of the differential forms \eqref{eq:05}.

\medbreak
Once relations \eqref{eq:06} are obtained, we substitute the differential forms \eqref{eq:05} into them
to arrive at systems of differential equations (often overdetermined)
for the unknown functions $\vartheta=\vartheta(x,t)$ and $\zeta=\zeta(u)$ that appear in \eqref{eq:02}.

\medbreak
\textit{Remark 1}.
Degenerate cases where one or more of the
differential forms $\Phi_n$ and/or $\Psi_n$ vanish in addition to
the linear relations \eqref{eq:06} must be treated separately.

\medbreak
\textit{Remark 2}.
The main ideas of the direct method of functional separation of variables based on transformation \eqref{eq:02} were expressed in the brief note \cite{pol2020},
where four exact solutions of a generalized porous medium equation with a nonlinear source were obtained.
The present paper demonstrates the effectiveness of this method by constructing a large number of solutions (more than 30 solutions have been obtained in total) to
a nonlinear diffusion-type equation involving several arbitrary functions.
In addition, it will be shown that the direct method is more efficient than indirect methods.

\medbreak
\textit{Remark 3}.
Bilinear functional-differential equations that are similar in appearance
to \eqref{eq:04}--\eqref{eq:05} arise when one searches for exact solutions to nonlinear equations of mathematical physics using the methods
of generalized and functional separation of variables with a priori given solution structure.
However, there is a fundamental difference in this case:
the differential forms $\Phi_n$ and $\Psi_n$ in \eqref{eq:05}
depend, in view of transformation \eqref{eq:02}, on the same independent variables $x$ and $t$, whereas when the methods of generalized and functional separation of variables
\cite{pol2005,pol2012,pol2019d,pol2019e,pol2019f} (see also \cite{bir1960}) are used,
the differential forms depend on different independent variables.
This circumstance significantly expands the possibilities of constructing exact solutions
by switching to equivalent equations
(see Section \ref{ss:2.3} for details).

\medbreak
\textit{Remark 4}.
Instead of transformation \eqref{eq:02}, we can use the transformation $\vartheta=Z(u)$, which leads to slightly more complex equations.
The method for constructing functional separable solutions described above is more convenient and is based on a substantial generalization
of traveling wave type solutions of various classes of nonlinear PDEs.
To illustrate this, consider the nonlinear heat equation
\begin{align}
u_{t}=[f(u)u_x]_x. \label{eq:004}
\end{align}
For arbitrary $f(u)$, equation \eqref{eq:004} admits the traveling wave solution
\begin{align}
u=u(z),\quad \ z=\kappa x+\lambda t, \label{eq:005}
\end{align}
where $\kappa$ and $\lambda$ \arbs.
Substituting \eqref{eq:005} in \eqref{eq:004} yields the ODE $\lambda u'_z=\kappa^2[f(u)u'_z]'_z$,
the integration of which gives its solution in implicit form
\begin{align}
\kappa x+\lambda t+C_1=\kappa^2\int \frac{f(u)\,du}{\lambda u+C_2},\label{eq:006}
\end{align}
where $C_1$ and $C_2$ \arbs.
On the left-hand side of \eqref{eq:006}, $z$ has been replaced with the original variables using \eqref{eq:005}.

The representation of the solution in the form \eqref{eq:02} is an essential generalization of the traveling wave solution \eqref{eq:006}, which is carried out as follows:
$$
\kappa x+\lambda t+C_1 \ \Longrightarrow \ \vartheta(x,t),\quad \
\frac{\kappa^2f(u)}{\lambda u+C_2} \ \Longrightarrow \ \zeta(u).
$$

\subsection{Some formulas allowing the satisfaction of relation (\ref{eq:04}) identically}\label{ss:2.2}

1.\enspace
For any $N$, equality \eqref{eq:04} can be satisfied if all $\Phi_i$ but one are put proportional to a selected
element $\Phi_j$ ($j\not=i$).
As a result, we get
\begin{equation}
\begin{aligned}
\Phi_i&=-A_i\Phi_j,\quad \ \ i=1,\dots,j-1,j+1,\dots,N;\\
\Psi_j&=A_1\Psi_1+\cdots+A_{j-1}\Psi_j+A_{j+1}\Psi_{j+1}+\cdots+A_N\Psi_N,
\end{aligned}
\label{eq:06**}
\end{equation}
where $A_i$ \arbs. In formulas \eqref{eq:06**},
the symbols can be swapped,
$\Phi_n \rightleftarrows \Psi_n$.

2.\enspace
For even $N$, equality \eqref{eq:04} is satisfied if $N/2$ individual pairwise sums $\Phi_i\Psi_i+\Phi_j\Psi_j$ vanish.
In this case, we have the relations
$$
\Phi_i-A_{ij}\Phi_j=0,\quad \ A_{ij}\Psi_i+\Psi_j=0\quad \ (i\not=j),
$$
where $A_{ij}$ \arbs\ and the indices $i$ and $j$ together take all values from 1 to $N$.

3.\enspace
For $N\ge 3$, equality \eqref{eq:04} is also satisfied identically
if we choose the linear relations
\begin{align}
\Phi_m-A_m\Phi_{N-1}-B_m\Phi_N&=0,\quad m=1,\,2,\,\dots,\,N-2;\notag\\
\Psi_{N-1}+A_1\Psi_1+\cdots+A_{N-2}\Psi_{N-2}&=0,\label{eq:06*}\\
\Psi_{N}+B_1\Psi_1+\cdots+B_{N-2}\Psi_{N-2}&=0,\notag
\end{align}
where $A_i$ and $B_i$ \arbs.
In formulas \eqref{eq:06*},
the symbols can be swapped, $\Phi_n \rightleftarrows \Psi_n$, or
simultaneous pairwise substitutions $\Phi_i \rightleftarrows \Phi_j$ and $\Psi_i \rightleftarrows \Psi_j$ can be made.

To construct more complex linear combinations of the form \eqref{eq:06} that would identically satisfy the bilinear
relation \eqref{eq:04}
for any $N$, one can use the formulas for the coefficients $\alpha_{ni}$ and $\beta_{nj}$ given in
the books \cite{pol2005,pol2012} (in sections devoted to generalized separation of variables).

\subsection{Possible generalizations based on the use of equivalent equations}\label{ss:2.3}

Other exact solutions of equation \eqref{eq:01} can be obtained if, instead of \eqref{eq:04}--\eqref{eq:05},
we consider equivalent differential equations
that reduce to \eqref{eq:04}--\eqref{eq:05} on the set of functions satisfying relation \eqref{eq:02}.
Indicated below are two classes of such equations, which will be used later in Section \ref{ss:3.3}.

1.\enspace One can use equations of the form
\begin{equation}
\sum^N_{n=1}\widetilde\Phi_n\widetilde\Psi_n=0,\quad \ \widetilde\Phi_n=\Phi_n\eta_n(\vartheta),\quad \widetilde\Psi_n=\Psi_n/\eta_n(Z),\quad Z=\int \zeta(u)\,du,
\label{eq:04*}
\end{equation}
which preserve the bilinear structure and, by virtue of \eqref{eq:02} (i.e.,  $\vartheta=Z$), are equivalent to equation \eqref{eq:04}--\eqref{eq:05}
for any functions $\eta_n(\vartheta)$.

2.\enspace One can use equations of the form
\begin{equation}
G(x,t,u,\vartheta)-G(x,t,u,Z)+
\sum^N_{n=1}\Phi_n\Psi_n=0,
\label{eq:04**}
\end{equation}
which for any functions $G(x,t,u,\vartheta)$ are equivalent to equation \eqref{eq:04}--\eqref{eq:05}.
Further, in Section \ref{ss:3.3}, specific examples of using equations of the form \eqref{eq:04**} for
$$
G(x,t,u,\vartheta)=\lambda(x,t,u)\vartheta
$$
will be given.
The functions $G$ and $\lambda$ can explicitly depend on $\vartheta$ and $\zeta$ (and their derivatives) and the functional coefficients of the original PDE
(which suggests implicit dependence on the original variables $x$, $t$, and $u$).

In the generic case, applying the splitting principle to equations \eqref{eq:04*} and \eqref{eq:04**} will lead
to other exact solutions of the original equation (1) than applying this principle to equation \eqref{eq:04}.

\medbreak
\textit{Remark 5}.
Further generalizations are also possible. In particular, the sum $\sum^N_{n=1}\Phi_n\Psi_n$ in \eqref{eq:04**}
can be replaced with $\sum^N_{n=1}\widetilde\Phi_n\widetilde\Psi_n$, where the tilde quantities are defined in \eqref{eq:04*}.
The functions $G(x,t,u,\vartheta)$ and $G(x,t,u,Z)$ can be multiplied by $\eta_{N+1}(\vartheta)/\eta_{N+1}(Z)$ and $\eta_{N+2}(\vartheta)/\eta_{N+2}(Z)$ respectively.

\section{Exact solutions of nonlinear equations of reaction-diffusion type}\label{s:3}

\subsection{The class of equations under consideration. Reduction to the bilinear form}\label{ss:3.1}

We consider a wide class of nonlinear diffusion equations,
\begin{align}
u_{t}=[a(x)f(u)u_x]_x+b(x)g(u)u_x+c(x)h(u), \label{eq:07}
\end{align}
which contain reaction and convective terms with variable coefficients.

Note that the exact solutions of some simpler equations belonging to class \eqref{eq:07} can be found, for example, in
\cite{doy1998,est2002,pol2005,gal2007,jia2008,pol2012,pol2019d,pol2019a,
pol2019b,nuc1992,kap2003,dor1982,kudr1993,gal1994,gan1999,pop2004,ivan2006,van2007,ivan2008,van2009,van2012,cher2013,bra2015,cher2018,bra2019}.

Using the method described in Section~\ref{s:2}, we further obtain a number of new exact solutions to equations of the form \eqref{eq:07},
in which at least two functional coefficients $a(x)$ and $f(u)$ are given arbitrarily (and the others are expressed through them).
Below,  for brevity, the arguments of the functions included in transformation \eqref{eq:02} and equation \eqref{eq:07} will often be omitted.

Having made the transformation \eqref{eq:02}, we substitute the derivatives \eqref{eq:03} in \eqref{eq:07}. After simple rearrangements we get
\begin{align}
-\vartheta_{t}+(a\vartheta_x)_xf+a\vartheta_x^2\Bl(\frac f\zeta \Br)^{\!\prime}_{\!u}+b\vartheta_x g+ch\zeta=0. \label{eq:08}
\end{align}

For $\zeta =1$, equation \eqref{eq:08} coincides with the original equation \eqref{eq:07}, where $u=\vartheta$.
Therefore, at this stage, no solutions are lost.

We introduce the following notation:
\begin{equation}
\begin{aligned}
\Phi_1&=-\vartheta_{t}, && \Phi_2=(a\vartheta_x)_x, && \Phi_3=a\vartheta_x^2, && \Phi_4=b\vartheta_x, &&\Phi_5=c;\\
\Psi_1&=1, &&  \Psi_2=f, && \Psi_3=(f/\zeta )^{\prime}_u, && \Psi_4=g, &&\Psi_5=h\zeta.
\end{aligned}
\label{eq:09}
\end{equation}
As a result, equation \eqref{eq:08} can be represented in the bilinear form \eqref{eq:04} with $N=5$:
\begin{equation}
\sum_{n=1}^5\Phi_n\Psi_n=0. \label{eq:10}
\end{equation}

We now turn to the construction of exact solutions of nonlinear equations of the form  \eqref{eq:07} based on relations \eqref{eq:09} and \eqref{eq:10}
using the approach described in Section \ref{ss:2.1}.

\subsection{Exact solutions obtained by analyzing equation (\ref{eq:08})}\label{ss:3.2}

\textbf{\textit{Solution 1}.}
Equation \eqref{eq:10} can be satisfied identically if we use the linear relations
\begin{equation}
\begin{aligned}
&\Phi_1=-\Phi_5,\quad \Phi_2=0,\quad k\Phi_3=-\Phi_4;\\
&\Psi_1=\Psi_5,\quad \Psi_3=k\Psi_4,
\end{aligned}
\label{eq:11}
\end{equation}
where $k$ \arb.
Substituting \eqref{eq:09} into \eqref{eq:11}, we arrive at the equations
\begin{equation}
\begin{aligned}
&\vartheta_t=c,\quad (a\vartheta_x)_x=0,\quad ka\vartheta_x^2=-b\vartheta_x;\\
&h\zeta =1,\quad (f/\zeta )^{\prime}_u=kg.
\end{aligned}
\label{eq:12}
\end{equation}

The solution of the overdetermined system consisting of the first three equations \eqref{eq:12} has the form
\begin{align}
\vartheta(x,t)=c_0t-\frac {b_0}k\int\frac{dx}{a(x)}+C_1,\quad b(x)=b_0,\quad c(x)=c_0, \label{eq:13}
\end{align}
where $a(x)$ \arbf\ and $b_0$, $c_0$, and $C_1$ \arbs.
The solution of the system consisting of the last two equations in \eqref{eq:12} can be written as follows:
\begin{align}
h=\frac{kG(u)+C_2}{f},\quad \zeta =\frac f{kG(u)+C_2},\quad G(u)=\int g(u)\,du, \label{eq:14}
\end{align}
where $f(u)$ and $g(u)$ \arbfs.
From formulas \eqref{eq:13} and \eqref{eq:14} for $b_0=c_0=1$ we obtain the equation
\begin{align}
u_{t}=[a(x)f(u)u_x]_x+g(u)u_x+\frac{kG(u)+C_2}{f(u)}, \label{eq:15}
\end{align}
which admits a generalized traveling wave solution in the implicit form
\begin{align}
\int\frac {f(u)\,du}{kG(u)+C_2}=t-\frac {1}k\int\frac{dx}{a(x)}+C_1.
\label{eq:16}
\end{align}

Note that equation \eqref{eq:15} contains three arbitrary functions $a(x)$, $f(u)$, and $g(u)$ and two arbitrary constants
$C_2$ and $k$.

\textbf{\textit{Solution 2}.}
Equation \eqref{eq:10} can be satisfied if we take
\begin{equation}
\begin{aligned}
&\Phi_1=-\Phi_4,\quad \Phi_2=0,\quad k\Phi_3=-\Phi_5;\\
&\Psi_1=\Psi_4,\quad \Psi_3=k\Psi_5,
\end{aligned}
\label{ee:700}
\end{equation}
where $k$ \arb.
Substituting \eqref{eq:09} into \eqref{ee:700}, we arrive at the equations
\begin{equation}
\begin{aligned}
&\vartheta_t=b\vartheta_x,\quad (a\vartheta_x)_x=0,\quad ka\vartheta_x^2=-c;\\
&g=1,\quad (f/\zeta )^{\prime}_u=kh\zeta.
\end{aligned}
\label{ee:701}
\end{equation}
The solutions of the first three equations \eqref{ee:701} are
\begin{equation}
\vartheta(x,t)=\lambda t+C_1\int\frac{dx}{a(x)}+C_2,\quad
b(x)=\frac{\lambda}{C_1}a(x),\quad c(x)=-\frac{kC_1^2}{a(x)},
\label{ee:702}
\end{equation}
where $a(x)$ \arbf\ and $C_1$, $C_2$, and $\lambda$ \arbs. The last two equations \eqref{ee:701} give two functions
\begin{equation}
g(u)=1,\quad \zeta(u)=\pm f(u)\BL(2k\int f(u)h(u)\,du+C_3\BR)^{\!-1/2},
\label{ee:703}
\end{equation}
where $f=f(u)$ and $h=h(u)$ \arbfs\ and $C_3$ \arb.

Setting $C_1=1$ in \eqref{ee:702} and \eqref{ee:703}, we obtain the equation
\begin{equation}
u_t=[a(x)f(u)u_x]_x+\lambda a(x)u_x-\frac{k}{a(x)}h(u),
\label{ee:704}
\end{equation}
where $a(x)$, $f(u)$, and $h(u)$ \arbfs, while $k$ and $\lambda$ \arbs. This equation admits two exact solutions
\begin{equation}
\pm\int f(u)\BL(2k\int f(u)h(u)\,du+C_3\BR)^{\!-1/2}du=\lambda t+\int\frac{dx}{a(x)}+C_2,
\label{ee:705}
\end{equation}
where $C_2$ and $C_3$ \arbs.

\textbf{\textit{Solution 3}.}
Equation \eqref{eq:10} can be satisfied by setting
\begin{equation}
\begin{aligned}
&\Phi_1=-k_1\Phi_5,\quad \Phi_2=-k_2\Phi_5,\quad \Phi_4=-k_3\Phi_5;\\
&\Psi_3=0,\quad \Psi_5=k_1\Psi_1+k_2\Psi_2+k_3\Psi_4,
\end{aligned}
\label{eq:19a}
\end{equation}
where $k_1$, $k_2$, and $k_3$ \arbs.
Substituting \eqref{eq:09} in \eqref{eq:19a}, we get
\begin{equation}
\begin{aligned}
&\vartheta_t=k_1c,\quad (a\vartheta_x)_x=-k_2c,\quad b\vartheta_x=-k_3c;\\
&(f/\zeta )^{\prime}_u=0,\quad h\zeta =k_1+k_2f+k_3g.
\end{aligned}
\label{eq:20a}
\end{equation}
The solution of the overdetermined system consisting of the first three equations \eqref{eq:20a} can be represented as
\begin{equation}
\begin{aligned}
\vartheta(x,t)&=c_0k_1t-c_0k_2\int\frac{x\,dx}{a(x)}-C_1\int\frac{dx}{a(x)}+C_2,\\
b(x)&=\frac{c_0k_3a(x)}{c_0k_2x+C_1},\quad c(x)=c_0,
\end{aligned}
\label{eq:21a}
\end{equation}
where $a(x)$ \arbf, while $c_0$, $C_1$, and $C_2$ \arbs.
From the last two equations \eqref{eq:20a} we obtain
\begin{equation}
h=\frac {k_1}f+k_2+k_3\frac gf,\quad \zeta=f,
\label{eq:21ab}
\end{equation}
where $f=f(u)$ and $g=g(u)$ \arbfs.

For $c_0=k_3=1$, formulas \eqref{eq:21a} and \eqref{eq:21ab} lead to the equation
\begin{align*}
u_{t}=[a(x)f(u)u_x]_x+\frac{a(x)}{k_2x+C_1}g(u)u_x+\frac{k_1+k_2f(u)+g(u)}{f(u)},
\end{align*}
which has the generalized traveling wave solution
\begin{align*}
\int f(u)\,du=k_1t-k_2\int\frac{x\,dx}{a(x)}-C_1\int\frac{dx}{a(x)}+C_2.
\end{align*}

\textbf{\textit{Solution 4}.}
Equation \eqref{eq:10} holds if we set
\begin{equation}
\begin{aligned}
&\Phi_1=-\Phi_5,\quad \Phi_2=-k\Phi_4;\\
&\Psi_1=\Psi_5,\quad k\Psi_2=\Psi_4,\quad \Psi_3=0,
\end{aligned}
\label{eq:19}
\end{equation}
where $k$ \arb.
Substituting \eqref{eq:09} into \eqref{eq:19} yields
\begin{equation}
\begin{aligned}
&\vartheta_t=c,\quad (a\vartheta_x)_x=-kb\vartheta_x;\\
&h\zeta =1,\quad kf=g,\quad (f/\zeta )^{\prime}_u=0.
\end{aligned}
\label{eq:20}
\end{equation}

The general solution of the overdetermined system consisting of the first two equations \eqref{eq:20} has the form
\begin{equation}
\begin{aligned}
\vartheta(x,t)&=c(x)t+s(x),\\
c(x)&=C_1\int\exp\BL(-k\int\frac ba\,dx\BR)\frac{dx}a+C_2,\\
s(x)&=C_3\int\exp\BL(-k\int\frac ba\,dx\BR)\frac{dx}a+C_4,
\end{aligned}
\label{eq:21}
\end{equation}
where $a=a(x)$ and $b=b(x)$ \arbfs, while $C_1$, $C_2$, $C_3$, and $C_4$ \arbs.
The solution of the system consisting of the last three equations \eqref{eq:20} is given by
\begin{align}
g=kf,\quad h=\frac 1{f},\quad \zeta =f.
\label{eq:22}
\end{align}
Given relations \eqref{eq:21} and \eqref{eq:22}, we obtain the equation
\begin{align}
u_{t}=[a(x)f(u)u_x]_x+kb(x)f(u)u_x+\frac{c(x)}{f(u)}, \label{eq:23}
\end{align}
which admits an exact solution in the implicit form
\begin{align}
\int f(u)\,du=c(x)t+s(x).
\label{eq:23a}
\end{align}
Here $a(x)$, $b(x)$, and $f(u)$ \arbfs,
and the functions $c(x)$ and $s(x)$ are defined in \eqref{eq:21}. In particular, for
$C_2=\lambda$, $C_1=0$, and $k=1$, we get the equation
\begin{align}
u_{t}=[a(x)f(u)u_x]_x+b(x)f(u)u_x+\frac{\lambda}{f(u)},\label{eq:23*}
\end{align}
which has the solution
\begin{align}
\int f(u)\,du=\lambda t+C_3\int\exp\BL(-\int\frac{b(x)}{a(x)}\,dx\BR)\frac{dx}{a(x)}+C_4.
\label{eq:23a*}
\end{align}

\textbf{\textit{Solution 5}.}
Equation \eqref{eq:10} can be satisfied by setting
\begin{equation}
\begin{aligned}
&\Phi_1+\Phi_2+\Phi_4=0,\quad \Phi_3=-k\Phi_5;\\
&\Psi_2=\Psi_1,\quad \Psi_4=\Psi_1,\quad k\Psi_3=\Psi_5,
\end{aligned}
\label{eq:400}
\end{equation}
where $k$ \arb. Substituting \eqref{eq:09} in \eqref{eq:400}, we get
\begin{equation}
\begin{aligned}
&{-\vartheta_{t}}+(a\vartheta_x)_x+b\vartheta_x=0,\quad a\vartheta_x^2=-kc;\\
&f=g=1,\quad k(f/\zeta)^{\!\prime}_{\!u}=h\zeta.
\end{aligned}
\label{eq:401}
\end{equation}
The first two equations \eqref{eq:401} admit the solution
\begin{align}
\vartheta(x,t)=\lambda t+\int r(x)\,dx+C_1,\quad b=\frac\lambda r-\frac{(ar)'_x}r,\quad c=-\frac{ar^2}k,
\label{eq:402}
\end{align}
where $a=a(x)$ and $r=r(x)$ \arbfs, while $\lambda$ and $C_1$ \arbs.
From the last equation \eqref{eq:401} we get $k\zeta^{-3}\zeta'_u=-h$, which gives two
solutions
\begin{align}
\zeta=\pm\BL(\frac 2k\int h\,du+C_2\BR)^{-1/2},
\label{eq:402}
\end{align}
where $h=h(u)$ \arbf\ and $C_2$ \arb.

\textbf{\textit{Solution 6}.}
Equation \eqref{eq:10} holds if we set
\begin{equation}
\begin{aligned}
&\Phi_1=\lambda\Phi_5,\quad \Phi_2=k_1\Phi_5,\quad \Phi_4=k_2\Phi_3;\\
&\lambda\Psi_1+k_1\Psi_2+\Psi_5=0,\quad \Psi_3=-k_2\Psi_4,
\end{aligned}
\label{eq:500}
\end{equation}
where $k_1$, $k_2$, and $\lambda$ \arbs.
Substituting \eqref{eq:09} into \eqref{eq:500}, we obtain
\begin{equation}
\begin{aligned}
&\vartheta_t=-\lambda c,\quad (a\vartheta_x)_x=k_1c,\quad b\vartheta_x=k_2a\vartheta_x^2;\\
&\lambda +k_1f+h\zeta=0,\quad (f/\zeta)^{\prime}_u=-k_2g.
\end{aligned}
\label{eq:501}
\end{equation}
The solution of the first three equations \eqref{eq:501} is expressed as
\begin{equation}
\begin{aligned}
&\vartheta(x,t)=-\lambda t+k_1\int\frac{x\,dx}{a(x)}+C_1\int\frac{dx}{a(x)}+C_2,\\
&b(x)=k_2(k_1x+C_1),\quad c(x)=1,
\end{aligned}
\label{eq:502}
\end{equation}
where $a(x)$ \arbf, while $C_1$ and $C_2$ \arbs.
The solution of the last two equations \eqref{eq:501} is given by
\begin{equation}
h=\frac{k_1f+\lambda}f\BL(k_2\int g\,du+C_3\BR),\quad \zeta=-f\BL(k_2\int g\,du+C_3\BR)^{\!-1},
\label{eq:503}
\end{equation}
where $f=f(u)$ and $g=g(u)$ \arbfs, while $C_3$ \arb.

Setting $k_1=k$ and $k_2=1$ in \eqref{eq:502} and \eqref{eq:503}, we arrive at the equation
$$
u_t=[a(x)f(u)u_x]_x+(kx+C_1)g(u)u_x+\frac{kf(u)+\lambda}{f(u)}G(u), \quad \ G(u)=\int g(u)\,du+C_3,
$$
where $a(x)$, $f(u)$, and $g(u)$ \arbfs, while $C_1$, $C_3$, $k$, and $\lambda$ \arbs.
This equation admits the exact solution in implicit form
$$
\int\frac {f(u)}{G(u)}\,du=\lambda t-k\int\frac{x\,dx}{a(x)}-C_1\int\frac{dx}{a(x)}-C_2.
$$

\textbf{\textit{Solution 7}.}
Equation \eqref{eq:10} can be satisfied by setting
\begin{equation}
\begin{aligned}
&\Phi_2=k_1\Phi_5,\quad \Phi_3=-k_2^2\Phi_1,\quad \Phi_4=-k_3\Phi_1;\\
&\Psi_5=-k_1\Psi_2,\quad \Psi_1-k_2^2\Psi_3-k_3\Psi_4=0,
\end{aligned}
\label{eq:504}
\end{equation}
where $k_1$, $k_2$, and $k_3$ \arbs.
Substituting \eqref{eq:09} in \eqref{eq:504} yields
\begin{equation}
\begin{aligned}
&(a\vartheta_x)_x=k_1c,\quad a\vartheta_x^2=k_2^2\vartheta_t,\quad b\vartheta_x=k_3\vartheta_t;\\
&h\zeta =-k_1f,\quad 1-k_2^2(f/\zeta )^{\prime}_u-k_3g=0.
\end{aligned}
\label{eq:505}
\end{equation}
The solutions of the first three equations \eqref{eq:505} can be represented as
\begin{equation}
\vartheta(x,t)=\lambda t+k_2\sqrt\lambda\int\frac{dx}{\sqrt{a}}+C_1,\quad b(x)=\frac{k_3}{k_2}\sqrt{\lambda a},\quad
c(x)=\frac{k_2\sqrt \lambda}{2k_1}\frac{a'_x}{\sqrt{a}},
\label{eq:506}
\end{equation}
where $a=a(x)$ \arb, while $C_1$ and $\lambda$ \arbs.
The solutions of the last two equations \eqref{eq:505} are given by
\begin{equation}
g=\frac 1{k_3}\BL(1+\frac{k_2^2}{k_1}h'_u\BR),\quad \zeta=-k_1\frac fh,
\label{eq:507}
\end{equation}
where $f=f(u)$ and $h=h(u)$ \arbfs.

Setting $k_1=k_3=1$, $k_2=1/\sqrt\lambda$, and $C_2=-C$ in \eqref{eq:506} and \eqref{eq:507}, we arrive at the equation
\begin{equation}
u_t=[a(x)f(u)u_x]_x+\sqrt{a(x)}[\lambda+h'_u(u)]u_x+\frac{1}2\frac{a'_x(x)}{\sqrt{a(x)}}h(u),
\label{eq:508}
\end{equation}
which contains three arbitrary functions $a(x)$, $f(u)$, and $h(u)$ and has the exact solution
\begin{equation}
\int\frac{f(u)}{h(u)}\,du=-\lambda t-\int\frac{dx}{\sqrt{a(x)}}+C.
\label{eq:508}
\end{equation}

\textbf{\textit{Solution 8}.}
Equation \eqref{eq:10} can be satisfied if we take
\begin{equation}
\begin{aligned}
&\Phi_1=-k_1\Phi_4,\quad \Phi_2=-k_2\Phi_4,\quad \Phi_3=-\Phi_5;\\
&\Psi_4=k_1\Psi_1+k_2\Psi_2,\quad \Psi_3=\Psi_5,
\end{aligned}
\label{eq:700}
\end{equation}
where $k_1$ and $k_2$ \arbs.
Substituting \eqref{eq:09} into \eqref{eq:700}, we arrive at the equations
\begin{equation}
\begin{aligned}
&\vartheta_t=k_1b\vartheta_x,\quad (a\vartheta_x)_x=-k_2b\vartheta_x,\quad a\vartheta_x^2=-c;\\
&g=k_1+k_2f,\quad (f/\zeta )^{\prime}_u=h\zeta.
\end{aligned}
\label{eq:701}
\end{equation}
The solutions of the first three equations \eqref{eq:701} are
\begin{equation}
\begin{aligned}
&\vartheta(x,t)=\lambda t-\frac{k_2\lambda}{k_1}\int\frac{x+C_1}{a(x)}\,dx+C_2,\\
&b(x)=-\frac{a(x)}{k_2(x+C_1)},\quad c(x)=-\frac{k_2^2\lambda^2(x+C_1)^2}{k_1^2a(x)},
\end{aligned}
\label{eq:702}
\end{equation}
where $a(x)$ \arbf, while $C_1$, $C_2$, and $\lambda$ \arbs. The last two equations \eqref{eq:701} give two solutions
\begin{equation}
g(u)=k_1+k_2f(u),\quad \zeta(u)=\pm f(u)\BL(2\int f(u)h(u)\,du+C_3\BR)^{\!-1/2},
\label{eq:703}
\end{equation}
where $f=f(u)$ and $h=h(u)$ \arbfs\ and $C_3$ \arb.

Setting $C_1=s$, $k_1=-1$, $k_2=k$, and $\lambda=k$ in \eqref{eq:702} and \eqref{eq:703}, we obtain the equation
\begin{equation}
u_t=[a(x)f(u)u_x]_x-\frac{a(x)}{x+s}[k+f(u)]u_x-\frac{(x+s)^2}{a(x)}h(u),
\label{eq:703a}
\end{equation}
where $a(x)$, $f(u)$, and $h(u)$ \arbfs, while $k$ and $s$ \arbs. This equation admits the exact solutions
\begin{equation}
\pm\int f(u)\BL(2\int f(u)h(u)\,du+C_3\BR)^{\!-1/2}du=kt-\int\frac{x+s}{a(x)}\,dx+C_2,
\label{eq:703b}
\end{equation}
where $C_2$ and $C_3$ \arbs.

In the special case $k=-1$, $f(u)=1$, and $s=0$, equation \eqref{eq:703a} is reduced to a simpler equation,
$$
u_t=[a(x)u_x]_x-\frac{x^2}{a(x)}h(u),
$$
which was considered in \cite{pol2019d}.
Setting $h(u)=0$, $C_3=0$, and $s=0$  in \eqref{eq:703a}, and
renaming $a(x)$ to $xa(x)$, we obtain the equation
$$
u_t=[xa(x)f(u)u_x]_x-a(x)[k+f(u)]u_x,
$$
whose solutions are
$$
\pm\int f(u)\,du=kt-\int\frac{dx}{a(x)}+C_2.
$$

\textbf{\textit{Solution 9}.}
Equation \eqref{eq:10} holds if we set
\begin{equation}
\begin{aligned}
&\Phi_1+\Phi_3+k_1\Phi_4+\Phi_5=0,\quad \Phi_2+k_2\Phi_4=0;\\
&\Psi_3=\Psi_1,\quad \Psi_4=k_1\Psi_1+k_2\Psi_2,\quad \Psi_5=\Psi_1,
\end{aligned}
\label{eq:800}
\end{equation}
where $k_1$ and $k_2$ \arbs.
Substituting \eqref{eq:09} into \eqref{eq:800}, we obtain the equations
\begin{equation}
\begin{aligned}
&{-\vartheta_t}+a\vartheta_x^2+k_1b\vartheta_x+c=0,\quad (a\vartheta_x)_x+k_2b\vartheta_x=0;\\
&(f/\zeta )^{\prime}_u=1,\quad g=k_1+k_2f,\quad h\zeta=1.
\end{aligned}
\label{eq:801}
\end{equation}

In the special case $k_1=k_2=0$, the solution of system \eqref{eq:801} leads to the equation
\begin{align}
u_t=[a(x)f(u)u_x]_x+\BL[\lambda-\frac{\beta^2}{a(x)}\BR]\frac u{f(u)},
\label{eq:801ab}
\end{align}
where $a(x)$ and $f(u)$ \arbfs, while $\beta$ and $\lambda$ \arbs. This equation admits two exact solutions
\begin{align}
\int\frac{f(u)}u\,du=\lambda t\pm \beta\int \frac{dx}{a(x)}+C_1.
\label{eq:801ac}
\end{align}

\textbf{\textit{Solution 10}.}
Equation \eqref{eq:10} can be satisfied if we take
\begin{equation}
\begin{aligned}
&\Phi_1=-k_1\Phi_5,\quad \Phi_2=-k_2\Phi_3,\quad \Phi_4=-k_3\Phi_5;\\
&\Psi_5=k_1\Psi_1+k_3\Psi_4,\quad \Psi_3=k_2\Psi_2,
\end{aligned}
\label{eq:900}
\end{equation}
where $k_1$, $k_2$, and $k_3$ \arbs.
Substituting \eqref{eq:09} into \eqref{eq:900}, we arrive at the equations
\begin{equation}
\begin{aligned}
&\vartheta_t=k_1c,\quad (a\vartheta_x)_x=-k_2a\vartheta_x^2,\quad b\vartheta_x=-k_3c;\\
&h\zeta=k_1+k_3g,\quad (f/\zeta )^{\prime}_u=k_2f.
\end{aligned}
\label{eq:901}
\end{equation}
The solutions of the first three equations \eqref{eq:901} are
\begin{equation}
\begin{aligned}
&\vartheta(x,t)=k_1t+\frac 1{k_2}\ln\BL(k_2\int\frac{dx}{a(x)}+C_1\BR)+C_2,\\
&b(x)=-k_3a(x)\BL(k_2\int\frac{dx}{a(x)}+C_1\BR),\quad c(x)=1,
\end{aligned}
\label{eq:902}
\end{equation}
where $a(x)$ \arbf, while $C_1$ and $C_2$ \arbs.
The solutions of the last two equations \eqref{eq:901} are given by
\begin{equation}
\begin{aligned}
&h(u)=\frac {k_1+k_3g(u)}{f(u)}\BL[k_2\int f(u)\,du+C_3\BR],\\
&\zeta(u)=f(u)\BL[k_2\int f(u)\,du+C_3\BR]^{\!-1},
\end{aligned}
\label{eq:903}
\end{equation}
where $f(u)$ and $g(u)$ \arbfs\ and $C_3$ \arb.

In particular, setting $a(x)=x^n$, $C_1=C_2=0$, $C_3=m$, $k_1=k$, $k_2=1-n$, and $k_3=1$ in \eqref{eq:902}--\eqref{eq:903},
we obtain the equation
$$
u_t=[x^nf(u)u_x]_x-xg(u)u_x+\frac {k+g(u)}{f(u)}\BL[(1-n)\int f(u)\,du+m\BR].
$$

\textbf{\textit{Solution 11}.}
Equation \eqref{eq:10} can be satisfied if we use the relations
\begin{equation}
\begin{aligned}
&\Phi_3=\Phi_1,\quad \Phi_4=k_1\Phi_1+k_2\Phi_2,\quad \Phi_5=\Phi_1;\\
&\Psi_1+\Psi_3+k_1\Psi_4+\Psi_5=0,\quad \Psi_2+k_2\Psi_4=0,
\end{aligned}
\label{eq:800*}
\end{equation}
where $k_1$ and $k_2$ \arbs.
Substituting \eqref{eq:09} in \eqref{eq:800*} yields
\begin{equation}
\begin{aligned}
&a\vartheta_x^2=-\vartheta_t,\quad b\vartheta_x=-k_1\vartheta_t+k_2(a\vartheta_x)_x,\quad c=-\vartheta_t;\\
&1+(f/\zeta )^{\prime}_u+k_1g+h\zeta=0,\quad f+k_2g=0.
\end{aligned}
\label{eq:801*}
\end{equation}

The first three equations \eqref{eq:801*} admit two solutions, which are given by
\begin{align}
\vartheta(x,t)=-t\pm\int\frac{dx}{\sqrt{a}}+C_1, \ \ \ b(x)=\pm k_1\sqrt{a}+\frac12k_2a'_x, \ \ \ c(x)=1,
\label{eq:801**}
\end{align}
where $a=a(x)$ \arbf\ and $C_1$ \arb \
(in both formulas, the upper or lower signs are taken simultaneously).
From the last equation \eqref{eq:801*} we get $g=-f/k_2$; then the penultimate equation,
which serves to determine the function $\zeta$, is converted to the Abel equation of the second kind
\begin{align}
\xi\xi'_u+\BL(1-\frac{k_1}{k_2}f\BR)\xi+fh=0,\quad \ \zeta=f/\xi.
\label{eq:801**a}
\end{align}

Setting $k_1=\pm k$ and $k_2=1$ in \eqref{eq:801**} and \eqref{eq:801**a}, we obtain the equation
$$
u_t=[a(x)f(u)u_x]_x-[k\sqrt{a(x)}+\tfrac12a'_x(x)]f(u)u_x+h(u),
$$
which has two exact solutions that can be represented in implicit form
\begin{align}
\int\frac{f(u)}{\xi(u)}\,du=-t\pm\int\frac{dx}{\sqrt{a(x)}}+C_1,
\label{eq:1200}
\end{align}
where the function $\xi=\xi(u)$ is described by the Abel equation
$$
\xi\xi'_u+[1\mp kf(u)]\xi+f(u)h(u)=0.
$$
Exact solutions of the Abel equations for various functions $f(u)$ and $h(u)$ can be found in \cite{polzai2018}.

\textbf{\textit{Solution 12}.}
We set $a=b=c=1$ in \eqref{eq:08} and then make the substitution
\begin{align}
\vartheta=\bar\vartheta+\alpha x+\beta t,
\label{eq:200}
\end{align}
where $\alpha$ and $\beta$ are free parameters, to obtain
\begin{align}
-\bar\vartheta_{t}+\bar\vartheta_{xx}f+(\bar\vartheta_x+\alpha)^2\Bl(\frac f\zeta \Br)^{\!\prime}_{\!u}+\bar\vartheta_x g-\beta+\alpha g+h\zeta=0.
\label{eq:201}
\end{align}

Below we give three solutions of equation \eqref{eq:201}, which lead to different solutions of the original PDE \eqref{eq:07}.

1.\enspace A particular solution to equation \eqref{eq:201} is sought in the form
\begin{align}
\bar\vartheta=C_1e^{\lambda t+\gamma x}+C_2,\quad \zeta=f,
\label{eq:202}
\end{align}
where $C_1$ and $C_2$ \arbs.
We get
$$
C_1(-\lambda+\gamma^2f+\gamma g)e^{\lambda t+\gamma x}-\beta+\alpha g+h\zeta=0,
$$
which leads to the defining system of equations
\begin{equation}
\begin{aligned}
-\lambda+\gamma^2f+\gamma g=0,\quad \
-\beta+\alpha g+h\zeta=0.
\end{aligned}
\label{eq:203}
\end{equation}
By virtue of the second equality \eqref{eq:202}, the solutions of these equations are
\begin{align}
g=\frac\lambda\gamma-\gamma f,\quad h=\alpha\gamma+\BL(\beta-\frac{\alpha\lambda}\gamma\BR)\frac 1f,\quad \zeta=f.
\label{eq:204}
\end{align}
Thus, we arrive at the equation
\begin{align}
u_{t}=[f(u)u_x]_x+\BL[\frac\lambda\gamma-\gamma f(u)\BR]u_x
+\alpha\gamma+\BL(\beta-\frac{\alpha\lambda}\gamma\BR)\frac 1{f(u)},
\label{eq:205}
\end{align}
which depends on an arbitrary function $f=f(u)$ and admits the exact solution in implicit form
\begin{align}
\int f(u)\,du=\alpha x+\beta t+C_1e^{\lambda t+\gamma x}+C_2.
\label{eq:206}
\end{align}
Setting $\lambda/\gamma=\sigma$, $\beta-(\alpha\lambda/\gamma)=\mu$, and $\alpha\gamma=\varepsilon$ in \eqref{eq:205} and \eqref{eq:206},
we obtain the more compact equation
\begin{align}
u_{t}=[f(u)u_x]_x+[\sigma-\gamma f(u)]u_x
+\varepsilon+\frac \mu{f(u)},
\label{eq:205*}
\end{align}
which has the exact solution
\begin{align}
\int f(u)\,du=\frac\varepsilon\gamma x+\Bl(\mu+\frac{\varepsilon\sigma}\gamma\Br)t+
C_1e^{\gamma\sigma t+\gamma x}+C_2.
\label{eq:206*}
\end{align}

2.\enspace
For $g\equiv 0$, equation \eqref{eq:201} has the steady-state particular solution
\begin{align}
\bar\vartheta=-kx^2+C,\quad h=\frac\beta f+2k,\quad\zeta=f,
\label{eq:207}
\end{align}
where $f=f(u)$ \arbf, while $C$ and $k$ \arbs. This leads to the PDE \cite{pol2012}
\begin{align}
u_{t}=[f(u)u_x]_x+2k+\frac \beta{f(u)}.
\label{eq:208}
\end{align}
This equation admits a solution in the implicit form $\int f(u)\,du=-kx^2+\alpha x+\beta t+C$.

3.\enspace
For $g\equiv 0$ and $\alpha=0$, equation \eqref{eq:201} has another steady-state particular solution
\begin{align}
\bar\vartheta=\ln(C_1x+C_2),\quad h=\beta\frac Ff,\quad\zeta=\frac fF,\quad F=\int f(u)\,du,
\label{eq:211}
\end{align}
which also leads to the equation considered in \cite{pol2012}.

\textbf{\textit{Solution 13}.}
In \eqref{eq:08} we set $\zeta=f$ and then make the transformation
\begin{align}
\vartheta=\bar\vartheta+\beta t+k\int\frac{dx}{a(x)},
\label{eq:250}
\end{align}
where $\beta$ and $k$ are free parameters, to obtain
\begin{align}
-\bar\vartheta_{t}+(a\bar\vartheta_x)_xf+b\bar\vartheta_x g-\beta+k\frac ba g+cfh=0.
\label{eq:251}
\end{align}

We are looking for a steady-state solution $\bar\vartheta=\bar\vartheta(x)$ of equation \eqref{eq:251}.
After the splitting procedure, we get the equations
\begin{align*}
&g=-\mu f+\lambda,\quad h=\gamma+(\sigma/f),\\
&(a\bar\vartheta^\prime_x)^\prime_x-\mu b\bar\vartheta^\prime_x+c\gamma-k\mu(b/a)=0,\\
&b\lambda\bar\vartheta^\prime_x g+\sigma -\beta+k\lambda(b/a)=0,
\end{align*}
where $\mu$, $\lambda$, $\gamma$, and $\sigma$ \arbs.
These equations admit the solution
\begin{equation}
\begin{aligned}
&\lambda=0,\quad \gamma=k\mu,\quad \sigma=\beta,\quad g(u)=-\mu f(u),\quad h(u)=k\mu+\frac{\beta}{f(u)},\\
&\bar\vartheta(x)=C_1\int\frac{e^{\mu x}}{a(x)}\,dx+C_2,\quad b(x)=a(x),\quad c(x)=1,\\
\end{aligned}
\label{eq:251*}
\end{equation}
where $C_1$, $C_2$, and $\mu$ \arbs.
Taking into account relation \eqref{eq:250}, we obtain the equation
$$
u_t=[a(x)f(u)u_x]_x-\mu a(x)f(u)u_x+\sigma +\frac{\beta}{f(u)},
$$
which admits the exact solution
$$
\int f(u)\,du=\beta t+\frac{\sigma}\mu\int\frac{dx}{a(x)}+C_1\int\frac{e^{\mu x}}{a(x)}\,dx+C_2.
$$

\textbf{\textit{Solution 14}.}
We seek a particular solution to equation \eqref{eq:251} as the product of functions with different arguments
\begin{align}
\bar\vartheta=e^{\lambda t}\xi(x).
\label{eq:252}
\end{align}
As a result, we arrive at the equations
\begin{equation}
\begin{aligned}
&{-}\lambda\xi+(a\xi'_x)'_xf+b\xi'_x g=0,\\
&{-}\beta+k\frac ba g+cfh=0.
\end{aligned}
\label{eq:253}
\end{equation}
For $g=\text{const}$, we obtain $f=\text{const}$ and $h=\text{const}$, which corresponds to a linear equation.
Therefore, we further assume that $g\not=\text{const}$.

The first equation \eqref{eq:253} is satisfied if we put
\begin{align}
(a\xi'_x)'_x-Ab\xi'_x=0,\quad Bb\xi'_x-\lambda\xi=0,\quad g=B-Af,
\label{eq:254}
\end{align}
where $A$ and $B$ \arbs \ ($A\not=0$).
The first two equations \eqref{eq:254} involve three functions $a=a(x)$, $b=b(x)$, and $\xi=\xi(x)$, one of which can be considered arbitrary.

Assuming that the function $\xi=\xi(x)$ in \eqref{eq:254} is given, we find that
\begin{align}
a=\frac 1{\xi'_x}\BL(\frac{A\lambda}B\int \xi\,dx+C_1\BR),\quad b=\frac{\lambda\xi}{B\xi'_x},
\label{eq:255}
\end{align}
If we assume that the function $b=b(x)$ is given, then the solutions of the first two equations \eqref{eq:254} can be written as
\begin{equation}
\begin{aligned}
a(x)&=b(x)\exp\BL(-\frac\lambda B\int\frac{dx}{b(x)}\BR)\BL[A\int \exp\BL(\frac\lambda B\int\frac{dx}{b(x)}\BR)dx+C_1\BR],\\
\xi(x)&=C_2\exp\BL(\frac\lambda B\int\frac{dx}{b(x)}\BR),
\end{aligned}
\label{eq:256}
\end{equation}
where $C_1$ and $C_2$ \arbs \ $(C_2\not=0)$.

In particular, for $B=1$ and $b(x)=1$, from \eqref{eq:256} we find that
$$
a(x)=\frac A{\lambda}+C_1e^{-\lambda x},\quad \xi(x)=C_2e^{\lambda x},
$$
and for $B=1$ and $b(x)=x$ we get
$$
a(x)=\frac A{\lambda +1}x^2+C_1x^{1-\lambda},\quad \xi(x)=C_2x^\lambda.
$$

The last equation \eqref{eq:253} can be satisfied in two cases, which are considered below.

1.\enspace
For $\beta=0$, the solution of the last equation \eqref{eq:253} is given by
\begin{align}
c(x)=k\frac {b(x)}{a(x)},\quad h(u)=A-\frac B{f(u)},
\label{eq:257}
\end{align}
in the derivation of which the last relation in \eqref{eq:254} was taken into account.
Thus, the equation
$$
u_t=[a(x)f(u)u_x]_x+b(x)[B-Af(u)]u_x+k\frac {b(x)}{a(x)}\BL[A-\frac B{f(u)}\BR],
$$
where $b(x)$ and $f(u)$ \arbfs,
and $a=a(x)$ is expressed via $b=b(x)$ by \eqref{eq:256}, admits the solution
$$
\int f(u)\,du=k\int\frac{dx}{a(x)}+C_2e^{\lambda t}\exp\BL(\frac\lambda B\int\frac{dx}{b(x)}\BR).
$$

2.\enspace
For $k=0$, the solution of the last equation \eqref{eq:253} is
\begin{align}
c(x)=1,\quad h(u)=\frac\beta{f(u)}.
\label{eq:257}
\end{align}
As a result, we obtain the equation
$$
u_t=[a(x)f(u)u_x]_x+b(x)[B-Af(u)]u_x+\frac \beta{f(u)},
$$
where $b(x)$ and $f(u)$ \arbfs, and $a=a(x)$ is expressed via $b=b(x)$ by \eqref{eq:256},
which has the solution
$$
\int f(u)\,du=\beta t+C_2e^{\lambda t}\exp\BL(\frac\lambda B\int\frac{dx}{b(x)}\BR).
$$

\textbf{\textit{Solution 15}.}
Equation \eqref{eq:10} can be satisfied if we take $\Phi_i$ $(i=1,\,2,\,3,\,4)$
proportional to $\Phi_5$. As a result, we get
\begin{equation}
\begin{aligned}
&\Phi_1=k_1\Phi_5,\quad \Phi_2=k_2\Phi_5,\quad \Phi_3=k_3\Phi_5,\quad\Phi_4=k_4\Phi_5,\\
&k_1\Psi_1+k_2\Psi_2+k_3\Psi_3+k_4\Psi_4+\Psi_5=0.
\end{aligned}
\label{eq:160}
\end{equation}
Substituting \eqref{eq:09} in \eqref{eq:160} yields
\begin{equation}
\begin{aligned}
&\vartheta_t=-k_1c,\quad (a\vartheta_x)_x=k_2c,\quad a\vartheta_x^2=k_3c,\quad b\vartheta_x=k_4c;\\
&k_1+k_2f+k_3(f/\zeta )^{\prime}_u+k_4g+h\zeta=0.
\end{aligned}
\label{eq:161}
\end{equation}

Consider two cases.

1.\enspace The simplest solution of the first four equations \eqref{eq:161},
$$
a(x)=b(x)=c(x)=1,\quad \theta(x,t)=-k_1t+k_4x+C_1,\quad k_2=0,\quad k_3=k_4^2,
$$
leads to a traveling wave solution of the original reaction-diffusion equation \eqref{eq:07} (this solution will not be discussed here).

2.\enspace
The first four equations \eqref{eq:161} also admit a different solution
\begin{equation}
\begin{aligned}
&a(x)=x^2,\quad b(x)=x,\quad c(x)=1,\\
&\vartheta(x,t)=-k_1t+k_2\ln x+C_1, \quad k_3=k_2^2,\quad k_4=k_2.\\
\end{aligned}
\label{eq:162}
\end{equation}
Setting $k=k_1$ and $k_2=1$ in \eqref{eq:162} and using the last equation in \eqref{eq:161}, we arrive at the reaction-diffusion type equation
\begin{align}
u_{t}=[x^2f(u)u_x]_x+xg(u)u_x+h(u),
\label{eq:163}
\end{align}
where
\begin{align}
h(u)=-\frac {\xi(u)}{f(u)}\bl[k+f(u)+g(u)+\xi^{\prime}_u(u)\br],\quad \ \xi(u)=\frac{f(u)}{\zeta(u)},
\label{eq:164}
\end{align}
and $f=f(u)$, $g=g(u)$, and $\xi=\xi(u)$ \arbfs. This equation admits the exact invariant solution
\begin{align}
\int \frac{f(u)}{\xi(u)}\,du=-kt+\ln x+C_1.
\label{eq:165}
\end{align}

Note that the invariant solution \eqref{eq:165} of equation \eqref{eq:163} can be obtained in the standard way in the form $u=U(z)$ with $z=-kt+\ln x$
(in this case, relation \eqref{eq:164} between $f$, $g$, $h$, and $\xi$ is not used). The function $U(z)$ is described by the ordinary differential equation
$$
[f(U)U^\prime_z]^\prime_z+[f(U)+g(U)+k]U^\prime_z+h(U)=0.
$$

\textbf{\textit{Solution 16}.}
Equation \eqref{eq:10} holds if we set
\begin{equation}
\begin{aligned}
&\Phi_1=k_1\Phi_4+k_2\Phi_5,\quad \Phi_2=-k_3\Phi_3;\\
&\Psi_3=k_3\Psi_2,\quad \Psi_4=-k_1\Psi_1,\quad \Psi_5=-k_2\Psi_1,
\end{aligned}
\label{ee:706}
\end{equation}
where $k_1$, $k_2$, and $k_3$ \arbs. Thus, we obtain the equations
\begin{equation}
\begin{aligned}
&\vartheta_{t}=-k_1b\vartheta_x-k_2c,\quad (a\vartheta_x)_x=-k_3a\vartheta_x^2;\\
&(f/\zeta )^{\prime}_u=k_3f,\quad g=-k_1,\quad h\zeta=-k_2.
\end{aligned}
\label{ee:707}
\end{equation}
The solutions of the first two equations \eqref{ee:707} are given by
\begin{equation}
\begin{aligned}
&\vartheta(x,t)=\lambda t+\frac 1{k_3}\ln\BL(k_3\int\frac{dx}{a(x)}+C_1\BR)+C_2,\\
&b(x)=-\frac{k_2c(x)+\lambda}{k_1}a(x)\BL(k_3\int\frac{dx}{a(x)}+C_1\BR),
\end{aligned}
\label{ee:708}
\end{equation}
where $c(x)$ and $c(x)$ \arbfs, while $C_1$ and $C_2$ \arbs.
From the last three equations \eqref{ee:707} we get
\begin{equation}
\begin{aligned}
&h(u)=-\frac{k_2}{f(u)}\BL(k_3\int f(u)\,du+C_3\BR),\quad \zeta(u)=f(u)\BL(k_3\int f(u)\,du+C_3\BR)^{\!-1}.
\end{aligned}
\label{ee:709}
\end{equation}
Substituting $C_1=C_3=0$ and $k_3=1$ in \eqref{ee:708} and \eqref{ee:709}, we obtain the equation
\begin{equation}
\begin{aligned}
u_t=[a(x)f(u)u_x]_x+a(x)[c(x)+\lambda]\BL(\int\frac{dx}{a(x)}\BR)u_x-\frac{c(x)}{f(u)}\int f(u),
\end{aligned}
\label{ee:710}
\end{equation}
which has the solution
\begin{equation}
\int f(u)\,du=C_4e^{\lambda t}\int\frac{dx}{a(x)},
\label{ee:711}
\end{equation}
where $C_4$ \arb. When deriving formula \eqref{ee:711}, the equality
$$
\int f\BL(\int f(u)\,du+C_3\BR)^{\!-1}du=\ln\BL(\int f(u)\,du+C_3\BR)+\text{const}
$$
was taken into account.
Note that the diffusion term of equation \eqref{ee:710} vanishes on solution \eqref{ee:711}, $[a(x)f(u)u_x]_x=0$.

\subsection{Exact solutions obtained by analyzing equivalent equations}\label{ss:3.3}

Now, using the considerations outlined in Section \ref{ss:2.3}, we will obtain some other exact solutions to equation \eqref{eq:01}.
To this end, instead of \eqref{eq:04}--\eqref{eq:05}, we consider equivalent differential equations that reduce to \eqref{eq:04}--\eqref{eq:05}
on the set of functions satisfying relation \eqref{eq:02}.

\textbf{\textit{Solution 17}.}
Let us return to the class of reaction-diffusion equations of the form \eqref{eq:07}.
Having made substitution \eqref{eq:02}, instead of equation \eqref{eq:08}, we consider the more complex equation
\begin{align}
-e^{\lambda\vartheta}e^{-\lambda Z}\vartheta_{t}+(a\vartheta_x)_xf+a\vartheta_x^2\Bl(\frac f\zeta \Br)^{\!\prime}_{\!u}+b\vartheta_x g+ch\zeta=0,\label{eq:25}
\end{align}
where $Z=\int \zeta \,du$ and $\lambda$ \arb.
Equations \eqref{eq:08} and \eqref{eq:25} are equivalent since, by virtue of transformation \eqref{eq:02}, the relation $\vartheta=Z$ holds.

Equation \eqref{eq:25} can be represented in the bilinear form \eqref{eq:10} where
\begin{equation}
\begin{aligned}
\Phi_1&=-e^{\lambda\vartheta}\vartheta_{t}, && \Phi_2=(a\vartheta_x)_x, && \Phi_3=a\vartheta_x^2, && \Phi_4=b\vartheta_x, &&\Phi_5=c;\\
\Psi_1&=e^{-\lambda Z}, &&  \Psi_2=f, && \Psi_3=(f/\zeta )^{\prime}_u, && \Psi_4=g, &&\Psi_5=h\zeta.
\end{aligned}
\label{eq:26}
\end{equation}
As previously, equation \eqref{eq:10} can be satisfied using relations \eqref{eq:11}. Substituting \eqref{eq:26} into \eqref{eq:11}, we arrive at the equations
\begin{equation}
\begin{aligned}
&e^{\lambda\vartheta}\vartheta_t=c,\quad (a\vartheta_x)_x=0,\quad ka\vartheta_x^2=-b\vartheta_x;\\
&h\zeta =e^{-\lambda Z},\quad (f/\zeta )^{\prime}_u=kg,
\end{aligned}
\label{eq:27}
\end{equation}
which for $\lambda=0$ coincide with \eqref{eq:12}.
The solution of the overdetermined system consisting of the first three equations \eqref{eq:27} has the form
\begin{equation}
\begin{aligned}
&\vartheta(x,t)=\frac 1\lambda\ln(t+C_2)-\frac{b_0}k\int\frac{dx}{a(x)}+C_1,\\
&b(x)=b_0,\quad c(x)=\frac 1\lambda\exp\BL(-\frac{b_0\lambda}k\int\frac{dx}{a(x)}+C_1\lambda\BR),
\end{aligned}
\label{eq:28}
\end{equation}
where $a(x)$ \arbf\ and $b_0$, $C_1$, $C_2$, $k$, and $\lambda$ \arbs.
The solution of the system consisting of the last two equations \eqref{eq:27} is written as
\begin{align}
\zeta(u) =\frac {f(u)}{kG(u)+C_2},\quad h(u)=\frac 1{\zeta(u)} \exp\BL(-\lambda\int \zeta(u)\,du\BR),\quad G(u)=\int g(u)\,du,
\label{eq:29}
\end{align}
where $f(u)$ and $g(u)$ \arbfs.

\textbf{\textit{Solution 18}.}
Equation \eqref{eq:10} can also be satisfied using relations \eqref{eq:19a}. Substituting \eqref{eq:26} into \eqref{eq:19a} yields
\begin{equation}
\begin{aligned}
&e^{\lambda\vartheta}\vartheta_t=k_1c,\quad (a\vartheta_x)_x=-k_2c,\quad b\vartheta_x=-k_3c;\\
&(f/\zeta )^{\prime}_u=0,\quad h\zeta =k_1e^{-\lambda Z}+k_2f+k_3g.
\end{aligned}
\label{eq:20a*}
\end{equation}
The solution of the overdetermined system consisting of the first three equations \eqref{eq:20a*} is
\begin{equation}
\begin{aligned}
&\vartheta(x,t)=\frac 1\lambda\ln[k_1(\lambda t+C_1)c(x)],\\
&a(x)=\frac {c(x)}{c'_x(x)}\BL(C_2-k_2\lambda\int c(x)\,dx\BR),\quad b(x)=-\frac {k_3\lambda c^2(x)}{c'_x(x)},
\end{aligned}
\label{eq:20a**}
\end{equation}
where $c(x)$ \arbf \ (other than a constant), while $C_1$, $C_2$, and $\lambda$ \arbs.
The solutions of the last two equations \eqref{eq:20a*} are expressed as
\begin{align}
\!\!h(u)=\frac 1{f(u)}\BL[k_1\exp\BL(\!-\lambda\!\int\! f(u)\,du\!\BR)+k_2f(u)+k_3g(u)\BR],\quad \zeta(u)=f(u),
\label{eq:20bb}
\end{align}
where $f=f(u)$ and $g=g(u)$ \arbfs.

\textbf{\textit{Solution 19}.}
As before, equation \eqref{eq:10} can be satisfied using relations \eqref{eq:19}. Substituting \eqref{eq:26} into \eqref{eq:19}, we get the equations
\begin{equation}
\begin{aligned}
&e^{\lambda\vartheta}\vartheta_t=c,\quad (a\vartheta_x)_x=-kb\vartheta_x;\\
&h\zeta =e^{-\lambda Z},\quad kf=g,\quad (f/\zeta )^{\prime}_u=0.
\end{aligned}
\label{eq:30}
\end{equation}
The solution of the overdetermined system consisting of the first two equations \eqref{eq:30} can be written as
\begin{equation}
\begin{aligned}
&\vartheta(x,t)=\frac 1\lambda\ln(t+C_1)+C_2\int\exp\BL(-k\int\frac{b}a\,dx\BR)\frac{dx}{a}+C_3,\\
&c(x)=\frac 1\lambda\exp\BL[C_2\lambda\int\exp\BL(-k\int\frac{b}a\,dx\BR)\frac{dx}{a}+C_3\lambda\BR],
\end{aligned}
\label{eq:31}
\end{equation}
where $a=a(x)$ and $b=b(x)$ \arbfs, while $C_1$, $C_2$, $C_3$, $k$, and $\lambda$ \arbs.
The solution to the system consisting of the last three equations \eqref{eq:30} is given by
\begin{align}
g=kf,\quad h=\frac 1{mf}\exp\BL(-m\lambda\int f\,du\BR),\quad \zeta =mf,
\label{eq:32}
\end{align}
where $m\not=0$ \arb.

\textbf{\textit{Solution 20}.}
Substituting \eqref{eq:26} into \eqref{eq:160}, we arrive at the equations
\begin{equation}
\begin{aligned}
&e^{\lambda\vartheta} \vartheta_t=-k_1c,\quad (a\vartheta_x)_x=k_2c,\quad a\vartheta_x^2=k_3c,\quad b\vartheta_x=k_4c;\\
&k_1e^{-\lambda Z}+k_2f+k_3(f/\zeta )^{\prime}_u+k_4g+h\zeta=0.
\end{aligned}
\label{eq:770}
\end{equation}

The first four equations of system \eqref{eq:770} admit a solution for the functional coefficients in exponential form:
\begin{equation}
\begin{aligned}
&a(x)=b(x)=c(x)=e^{\lambda x},\quad \vartheta(x,t)=\frac 1\lambda \ln t+x,\\
&k_1=-\frac 1\lambda,\quad k_2=\lambda,\quad k_3=k_4=1.
\end{aligned}
\label{eq:771}
\end{equation}
Using the last equation in \eqref{eq:770}, we obtain the reaction-diffusion type equation
\begin{align}
u_{t}=[e^{\lambda x}f(u)u_x]_x+e^{\lambda x}g(u)u_x+e^{\lambda x}h(u),
\label{eq:772}
\end{align}
where
\begin{align}
h(u)=-\frac 1\zeta\BL[-\frac 1\lambda e^{-\lambda Z}+\lambda f+\Bl(\frac f\zeta\Br)^{\prime}_u+g\BR],\quad \ Z=\int \zeta\,du,
\label{eq:773}
\end{align}
and $f=f(u)$, $g=g(u)$, and $\zeta=\zeta(u)$ \arbfs. Equation \eqref{eq:772} admits the exact invariant solution
\begin{align}
\int \zeta(u)\,du=\frac 1\lambda \ln t+x.
\label{eq:774}
\end{align}

Note that the invariant solution \eqref{eq:774} of equation \eqref{eq:772} can be represented in the standard form $u=U(z)$
with $z=\frac 1\lambda \ln t+x$ (in this case, relation \eqref{eq:773} linking $f$, $g$, $h$ and $\zeta$ is not used).
The function $U(z)$ is described by the ordinary differential equation
$$
\frac 1\lambda U^\prime_z=[e^{\lambda z}f(U)U^\prime_z]^\prime_z+e^{\lambda z}g(U)U^\prime_z+e^{\lambda z}h(U).
$$

\textbf{\textit{Solution 21}.}
The first four equations of system \eqref{eq:770} also admit a solution for power-law functional coefficients:
\begin{equation}
\begin{aligned}
&a(x)=x^n,\quad b(x)=x^{n-1},\quad c(x)=x^{n-2},\quad \vartheta(x,t)=\frac 1{n-2}\ln t+\ln x,\\
&\lambda=n-2,\quad k_1=-\frac 1{n-2},\quad k_2=n-1,\quad k_3=k_4=1.
\end{aligned}
\label{eq:775}
\end{equation}
Using the last equation in \eqref{eq:770}, we arrive at the reaction-diffusion type equation
\begin{align}
u_{t}=[x^nf(u)u_x]_x+x^{n-1}g(u)u_x+x^{n-2}h(u),
\label{eq:776}
\end{align}
where
\begin{align}
h(u)=-\frac 1\zeta\BL[-\frac 1{n-2}e^{-(n-2)Z}+(n-1)f+\Bl(\frac f\zeta\Br)^{\prime}_u+g\BR],\quad \ Z=\int \zeta\,du,
\label{eq:777}
\end{align}
and $f=f(u)$, $g=g(u)$, and $\zeta=\zeta(u)$ \arbfs. Equation \eqref{eq:776} admits the exact invariant solution
\begin{align}
\int \zeta(u)\,du=\frac 1{n-2}\ln t+\ln x.
\label{eq:778}
\end{align}

The self-similar solution \eqref{eq:778} of equation \eqref{eq:776} can be sought in the standard form $u=U(z)$ with $z=xt^{1/(n-2)}$
(in this case, relation \eqref{eq:777} linking $f$, $g$, $h$, and $\zeta$ is not used).
The function $U(z)$ is described by the ODE
$$
\frac 1{n-2}zU^\prime_z=[z^nf(U)U^\prime_z]^\prime_z+z^{n-1}g(U)U^\prime_z+z^{n-2}g(U).
$$

\textbf{\textit{Solution 22}.}
Equation \eqref{eq:10} can be satisfied if we take $\Psi_i$ $(i=1,\,3,\,4,\,5)$ proportional to $\Psi_2$. As a result, we get
\begin{equation}
\begin{aligned}
&\Psi_1=k_1\Psi_2,\quad \Psi_3=k_2\Psi_2,\quad \Psi_4=k_3\Psi_2,\quad\Psi_5=k_4\Psi_2,\\
&k_1\Phi_1+\Phi_2+k_2\Phi_3+k_3\Phi_4+k_4\Phi_5=0.
\end{aligned}
\label{eq:779}
\end{equation}
Substituting \eqref{eq:26} into \eqref{eq:779}, we obtain the equations
\begin{equation}
\begin{aligned}
&e^{-\lambda Z}=k_1f,\quad (f/\zeta )^{\prime}_u=k_2f,\quad g=k_3f,\quad h\zeta=k_4f,\\
&{-}k_1e^{\lambda\vartheta}\vartheta_{t}+(a\vartheta_x)_x+k_2a\vartheta_x^2+k_3b\vartheta_x+k_4c=0.
\end{aligned}
\label{eq:780}
\end{equation}

The first four equations of system \eqref{eq:780} admit a solution for exponential functional coefficients:
\begin{equation}
\begin{aligned}
&f(u)=g(u)=h(u)=e^{-\lambda u},\quad \zeta=1,\quad Z=u;\\
&k_1=k_3=k_4=1,\quad k_2=-\lambda.
\end{aligned}
\label{eq:781}
\end{equation}
In this case, we obtain the reaction-diffusion type equation
\begin{align}
u_{t}=[a(x)e^{\beta u}u_x]_x+b(x)e^{\beta u}u_x+c(x)e^{\beta u},\quad \ \lambda=-\beta,
\label{eq:782}
\end{align}
which has the exact solution with additive separation of variables
\begin{align}
u=-\frac 1\beta \ln t+\eta(x),
\label{eq:783}
\end{align}
with the function $\eta=\eta(x)$ described by the ordinary differential equation
\begin{align}
-\frac 1\beta=[a(x)e^{\beta\eta}\eta'_x]'_x+b(x)e^{\beta\eta}\eta'_x+c(x)e^{\beta\eta}.
\label{eq:784}
\end{align}
Equations \eqref{eq:782} and \eqref{eq:784} contain three arbitrary functions
$a(x)$, $b(x)$, and $c(x)$.

Note that equation \eqref{eq:784} reduces with the substitution $\xi=e^{\beta\eta}$ to the linear second-order ODE
$$
[a(x)\xi'_x]'_x+b(x)\xi'_x+\beta c(x)\xi+1=0.
$$

\textbf{\textit{Solution 23}.}
The first four equations of system \eqref{eq:780} also admit  a solution for the power-law functional coefficients
\begin{equation}
\begin{aligned}
&f(u)=u^n,\quad g(u)=u^n,\quad h(u)=u^{n+1},\quad \zeta(u)=1/u,\quad Z=\ln u,\\
&\lambda=-n,\quad k_1=k_3=k_4=1,\quad k_2=n+1.
\end{aligned}
\label{eq:785}
\end{equation}
In this case, the solution of the last equation in \eqref{eq:780} is determined by the formula $\vartheta=-(1/n)\ln t+\eta(x)$,
with the function $\eta=\eta(x)$ satisfying the ODE
\begin{align}
\frac 1n e^{-n\eta}+(a\eta_x')'_x+(n+1)a(\eta'_x)^2+b\eta'_x+c=0.
\label{eq:786}
\end{align}
As a result, we get the reaction-diffusion type equation
\begin{align}
u_{t}=[a(x)u^nu_x]_x+b(x)u^nu_x+c(x)u^{n+1},
\label{eq:787}
\end{align}
the exact solution of which can be represented as the product of functions with different arguments $u=t^{-1/n}\xi(x)$, with the function $\xi(x)=e^{\eta}$ described by ODE
$$
[a(x)\xi^n\xi'_x]'_x+b(x)\xi^n\xi'_x+c(x)\xi^{n+1}+\frac 1n\xi=0.
$$

\textbf{\textit{Solution 24}.}
Let us return to the class of reaction-diffusion equations
of the form \eqref{eq:07}. Having made the substitution \eqref{eq:02}, instead of equation \eqref{eq:08}, we consider the more complex equation
\begin{align}
-\vartheta_{t}+(a\vartheta_x)_xf+a\vartheta_x^2\Bl(\frac f\zeta \Br)^{\!\prime}_{\!u}+b\vartheta_x g+ch\zeta\frac{\vartheta}Z=0,\label{eq:100}
\end{align}
where $Z=\int \zeta \,du$.
Equations \eqref{eq:08} and \eqref{eq:100} are equivalent, because, by virtue
of transformation (6), the relation $\vartheta=Z$ holds.

Equation \eqref{eq:100} can be represented in bilinear form \eqref{eq:10}, where
\begin{equation}
\begin{aligned}
\Phi_1&=-\vartheta_{t}, && \Phi_2=(a\vartheta_x)_x, && \Phi_3=a\vartheta_x^2, && \Phi_4=b\vartheta_x, &&\Phi_5=c\vartheta;\\
\Psi_1&=1, &&  \Psi_2=f, && \Psi_3=(f/\zeta )^{\prime}_u, && \Psi_4=g, &&\Psi_5=h\zeta/Z.
\end{aligned}
\label{eq:101}
\end{equation}
Equation \eqref{eq:10} can be satisfied by using the relations \eqref{eq:19a}.
Substituting \eqref{eq:101} in \eqref{eq:19a}, we get
\begin{equation}
\begin{aligned}
&\vartheta_t=k_1c\vartheta,\quad (a\vartheta_x)_x=-k_2c\vartheta,\quad b\vartheta_x=-k_3c\vartheta;\\
&(f/\zeta)^{\prime}_u=0,\quad h\zeta/Z =k_1+k_2f+k_3g,
\end{aligned}
\label{eq:102}
\end{equation}
where $k_1$, $k_2$, and $k_3$ \arbs.
Let $a=a(x)$, $f=f(u)$, and $g=g(u)$ be arbitrary functions.
Then the solutions of equations \eqref{eq:102} are given by
\begin{equation}
\begin{aligned}
&b(x)=-\frac{k_3\lambda}{k_1}\frac{\omega}{\omega'_x},\quad c(x)=\frac\lambda{k_1}=\text{const},\quad
\vartheta(x,t)=e^{\lambda t}\omega(x),\\
&h=\frac 1f(k_1+k_2f+k_3g)F,\quad \zeta=f,\quad F=\int f\,du,
\end{aligned}
\label{eq:103}
\end{equation}
where $\lambda$ \arb, and the function $\omega=\omega(x)$ solves the linear second-order ODE
 $(a\omega'_x)'_x=-(k_2\lambda/k_1)\omega$.
 In the special case $a(x)=\text{const}$ and $k_3=0$, formulas \eqref{eq:103} lead to the nonlinear reaction-diffusion equation
 and its solution, which were considered in \cite{pol2012}.

\textbf{\textit{Solution 25}.}
Consider the special case
\begin{align}
a=b=c=1,\quad \zeta=f.
\label{eqq:001}
\end{align}
We look for a solution of equation \eqref{eq:100} under conditions \eqref{eqq:001} in the form
\begin{align}
\vartheta=(\gamma x+\delta)e^{\alpha x+\beta t}.
\label{eqq:002}
\end{align}
Substituting \eqref{eqq:002} into \eqref{eq:100} and taking into account \eqref{eqq:001}, we obtain
\begin{equation}
\begin{aligned}
&\gamma xe^{\alpha x+\beta t}\bl[-\beta+\alpha^2f+\alpha g+(f/F)h\br]\\
&\qquad+e^{\alpha x+\beta t}\bl[-\beta\delta+(\alpha^2\delta+2\alpha\gamma)f+(\alpha\delta+\gamma)g+\delta(f/F)h\br]=0,
\end{aligned}
\label{eqq:003}
\end{equation}
where $F=\int f\,du$.
Equating the expressions in square brackets in \eqref{eqq:003} with zero, we arrive at the equations
\begin{align*}
-\beta+\alpha^2f+\alpha g+(f/F)h&=0,\\
-\beta\delta+(\alpha^2\delta+2\alpha\gamma)f+(\alpha\delta+\gamma)g+\delta(f/F)h&=0.
\end{align*}
Solving these equations for $g$ and $h$, we get
$$
g=-2\alpha f,\quad \ h=\BL(\alpha^2+\frac\beta{f}\BR)F.
$$
As a result, we obtain the equation
\begin{align}
u_t=[f(u)u_x]_x-2\alpha f(u)u_x+\BL[\alpha^2+\frac{\beta}{f(u)}\BR]\int f(u)\,du,
\label{eqq:004}
\end{align}
which has the exact solution
\begin{align}
\int f(u)\,du=(\gamma x+\delta)e^{\alpha x+\beta t},
\label{eqq:005}
\end{align}
where $\gamma$ and $\delta$ \arbs.

\textbf{\textit{Solution 26}.}
We look for a solution to equation \eqref{eq:100} under conditions \eqref{eqq:001} in the form
\begin{align*}
\vartheta=Ae^{\alpha x+\beta t}+Be^{\gamma x+\delta t}.
\end{align*}
Omitting the intermediate calculations, we arrive at the equation
\begin{align}
u_t=[f(u)u_x]_x+\BL[\frac{\delta-\beta}{\gamma-\alpha}&-(\alpha+\gamma)f(u)\BR]u_x\notag\\
&+\BL[\alpha\gamma+\frac{\beta\gamma-\alpha\delta}{\gamma-\alpha}\frac{1}{f(u)}\BR]\int f(u)\,du,
\label{eqq:006}
\end{align}
which has the solution
\begin{align}
\int f(u)\,du=Ae^{\alpha x+\beta t}+Be^{\gamma x+\delta t},
\label{eqq:007}
\end{align}
where $A$ and $B$ \arbs.

\textit{Example 1}.
In the special case $\gamma=-\alpha$ and $\delta=\beta$, equation \eqref{eqq:006} simplifies and takes the form
\begin{align*}
u_t=[f(u)u_x]_x+\BL[-\alpha^2+\frac{\beta}{f(u)}\BR]\int f(u)\,du
\end{align*}
and its solution is written as
\begin{align*}
\int f(u)\,du=e^{\beta t}(Ae^{\alpha x}+Be^{-\alpha x}).
\end{align*}

\textbf{\textit{Solution 27}.}
Assuming that conditions \eqref{eqq:001} hold, we  look for a solution to equation \eqref{eq:100} in the form
\begin{align*}
\vartheta=Ae^{\alpha t}\sin(\beta x+\sigma t+\delta),
\end{align*}
where $A$ and $\delta$ \arbs.
After simple rearrangements, we obtain the equation
\begin{align}
u_t=[f(u)u_x]_x+\gamma u_x+\BL[\beta^2+\frac{\alpha}{f(u)}\BR]\int f(u)\,du
\label{eqq:008}
\end{align}
with $\gamma=\sigma/\beta$, which admits the exact solution
\begin{align}
\int f(u)\,du=Ae^{\alpha t}\sin(\beta x+\beta\gamma t+\delta).
\label{eqq:009}
\end{align}

\textit{Remark 6}.
In the case \eqref{eqq:001}, equation \eqref{eq:100} also admits a more complex solution of the form
$$
\vartheta=Ae^{\mu x+\alpha t}\sin(\beta x+\sigma t+\delta),
$$
which we do not consider here.

\textbf{\textit{Solution 28}.}
Instead of equation \eqref{eq:100}, we can look at the more complex equation
\begin{align}
-\frac{\vartheta^n}{Z^n}\vartheta_{t}+(a\vartheta_x)_xf+a\vartheta_x^2\Bl(\frac f\zeta \Br)^{\!\prime}_{\!u}+b\vartheta_x g+ch\zeta\frac{\vartheta}Z=0,\label{eq:100**}
\end{align}
where $Z=\int \zeta \,du$ and $n$ \arb.
Equations \eqref{eq:08} and \eqref{eq:100**} are equivalent, since, by virtue of transformation \eqref{eq:02}, the relation $\vartheta=Z$ holds.

Equation \eqref{eq:100**} can be represented in the bilinear form \eqref{eq:10} where
\begin{equation}
\begin{aligned}
\Phi_1&=-\vartheta^n\vartheta_{t}, && \Phi_2=(a\vartheta_x)_x, && \Phi_3=a\vartheta_x^2, && \Phi_4=b\vartheta_x, &&\Phi_5=c\vartheta;\\
\Psi_1&=Z^{-n}, &&  \Psi_2=f, && \Psi_3=(f/\zeta )^{\prime}_u, && \Psi_4=g, &&\Psi_5=h\zeta/Z.
\end{aligned}
\label{eq:101**}
\end{equation}
Equation \eqref{eq:10} can be satisfied by using the relations \eqref{eq:19a}. Substituting \eqref{eq:101**} in \eqref{eq:19a}, we get
\begin{equation}
\begin{aligned}
&\vartheta^n\vartheta_t=k_1c\vartheta,\quad (a\vartheta_x)_x=-k_2c\vartheta,\quad b\vartheta_x=-k_3c\vartheta;\\
&(f/\zeta)^{\prime}_u=0,\quad h\zeta/Z =k_1Z^{-n}+k_2f+k_3g,
\end{aligned}
\label{eq:102**}
\end{equation}
where $k_1$, $k_2$, and $k_3$ \arbs.
Let $a=a(x)$, $f=f(u)$, and $g=g(u)$ be arbitrary functions.
Then the solutions of equations \eqref{eq:102**} are expressed as
\begin{equation}
\begin{aligned}
&b(x)=-\frac{k_3}{k_1n}\frac{\omega^{n+1}}{\omega'_x},\quad c(x)=\frac{\omega^n}{k_1n},\quad
\vartheta(x,t)=t^{1/n}\omega(x),\\
&h=\frac 1f(k_1F^{-n}+k_2f+k_3g)F,\quad \zeta=f,\quad F=\int f\,du,
\end{aligned}
\label{eq:103**}
\end{equation}
where the function $\omega=\omega(x)$ is a solution of a second-order nonlinear ODE of the Emden--Fowler type:
\begin{align}
(a\omega'_x)'_x=-\frac{k_2}{k_1n}\omega^{n+1}.
\label{eq:103aa}
\end{align}

We set $k_3=k_1n$ and $k=k_2/(k_1n)$. From relations \eqref{eq:103**} it follows that the nonlinear reaction-diffusion type equation
\begin{align}
u_{t}=[a(x)f(u)u_x]_x-\frac{\omega^{n+1}}{\omega^\prime_x}g(u)u_x+\omega^n\frac {F(u)}{f(u)}\Bl[kf(u)+g(u)+\frac 1nF^{-n}(u)\Br],
\label{eq:105**}
\end{align}
where $f(u)$, $g(u)$, and $a(x)$ \arbfs, $k$ and $n$ \arbs, and $F(u)=\int f(u)\,du$,
admits the functional separable solution in implicit form
\begin{equation}
\int f(u)\,du=\omega(x)t^{1/n}.
\label{eq:106**}
\end{equation}
The function $\omega=\omega(x)$ in \eqref{eq:105**} and \eqref{eq:106**} is described by the nonlinear ordinary differential equation
\begin{equation}
[a(x)\omega'_x]'_x+k\omega^{n+1}=0.
\label{eq:107**}
\end{equation}

Note that for $n=-1$, the general solution of equation \eqref{eq:107**} is
\begin{equation*}
\xi=-k\int\frac {x\,dx}{a(x)}+C_1\int \frac{dx}{a(x)}+C_2,
\end{equation*}
where $C_1$ and $C_2$ \arbs.

\textit{Example 2}.
Substituting $a(x)=1$ and $k=0$ into \eqref{eq:105**}--\eqref{eq:107**}, we get the equation
\begin{equation}
u_t=[f(u)u_x]_x-x^{n+1}g(u)u_x+x^{n}\frac 1{f(u)}\Bl[g(u)F(u)+\frac 1nF^{1-n}(u)\Br],
\label{eq:108**}
\end{equation}
which admits the exact solution in implicit form \eqref{eq:106**}.
This solution is non-invariant and it is of a self-similar type;
when substituted into equation \eqref{eq:108**}, it causes the term $[f(u)u_x]_x$ to vanish.

\textbf{\textit{Solution 29}.}
We now consider the equation
\begin{align}
-\vartheta_{t}+\lambda\vartheta-\lambda Z +(a\vartheta_x)_xf+a\vartheta_x^2\Bl(\frac f\zeta \Br)^{\!\prime}_{\!u}+b\vartheta_x g+ch\zeta=0,
\label{eq:104}
\end{align}
where $\lambda=\text{const}$,
which, by virtue of \eqref{eq:02} ($\vartheta=Z$), is equivalent to equation \eqref{eq:08}.

Equation \eqref{eq:104} is invariant under the transformation
\begin{align}
\vartheta=\bar\vartheta+C_1e^{\lambda t},
\label{eq:105}
\end{align}
where $C_1$ \arb.

It is easy to verify that for constant $a$, $b$, and $c$, which without loss of generality can be set equal to 1, equation \eqref{eq:104} has the particular solution
\begin{align}
\bar\vartheta=C_2e^{\beta t-\mu x},\quad g=\mu f+\frac{\lambda-\beta}{\mu},\quad h=\frac{\lambda}{f}\int f\,du,\quad \zeta=f,
\label{eq:106}
\end{align}
where $f=f(u)$ \arbf, while $C_2$, $\beta$, and $\mu$ \arbs.
Given \eqref{eq:105}, we obtain the equation
\begin{align}
u_{t}=[f(u)u_x]_x+\BL[\mu f(u)+\frac{\lambda-\beta}{\mu}\BR]u_x+\frac{\lambda}{f(u)}\int f(u)\,du,
\label{eq:107}
\end{align}
which has the exact solution in the implicit form
\begin{align}
\int f(u)\,du=C_1e^{\lambda t}+C_2e^{\beta t-\mu x}.
\label{eq:108}
\end{align}
Setting $\beta=\lambda-\sigma\mu$, equation \eqref{eq:107} can be rewritten in the more compact form
$$
u_{t}=[f(u)u_x]_x+[\mu f(u)+\sigma]u_x+\frac{\lambda}{f(u)}\int f(u)\,du.
$$
In this case, its solution is
$\int f(u)\,du=C_1e^{\lambda t}+C_2e^{(\lambda-\sigma\mu)t-\mu x}$.

\textbf{\textit{Solution 30}.}
We look for a steady-state particular solution $\vartheta=\vartheta(x)$ of equation \eqref{eq:104}.
In this case, we have
\begin{equation}
\begin{aligned}
&a\vartheta_x^2=k_1\vartheta,\quad (a\vartheta_x)_x=k_2,\quad b\vartheta_x=k_3,\quad c=1;\\
&\lambda+k_1(f/\zeta)'_u=0,\quad -\lambda Z+k_2f+k_3g+h\zeta=0,
\end{aligned}
\label{eq:109}
\end{equation}
where $k_1$, $k_2$, and $k_3$ \arbs.
The solution of the first three equations \eqref{eq:109} with $k_1k_2\not=0$ can be represented as
\begin{equation}
\begin{aligned}
&a(x)=\frac 1{C_2k_1}(k_2x+C_3)^{2-(k_1/k_2)},\quad b(x)=\frac {k_3}{C_2k_1}(k_2x+C_3)^{1-(k_1/k_2)},\\
&\vartheta(x)=C_2(k_2x+C_3)^{k_1/k_2},
\end{aligned}
\label{eq:110}
\end{equation}
where $C_2$ and $C_3$ \arbs.
The solution to the system consisting of the last two equations \eqref{eq:109} is written as follows:
\begin{equation}
\zeta=-\frac{k_1}{\lambda}\frac f{u+C_4},\quad h=\frac{\lambda(u+C_4)}{k_1f}\BL(k_2f+k_3g+k_1\int\frac {f\,du}{u+C_4}\BR),
\label{eq:110a}
\end{equation}
where $f=f(u)$ and $g=g(u)$ \arbfs, $C_4$ \arb.

Substituting $C_2=1/k$, $C_3=C_4=0$, $k_1=k$, $k_2=k_3=1$, and $\lambda=k\sigma$ in \eqref{eq:110} and \eqref{eq:110a},  we arrive at the equation
\begin{equation}
u_t=[x^{2-k}f(u)u_x]_x+x^{1-k}g(u)u_x+\frac{\sigma u}{f(u)}\BL[f(u)+g(u)+k\int\frac{f(u)}u\,du\BR].
\label{eq:111}
\end{equation}
For $k\not=0$, this equation admits the exact solution
\begin{equation}
\int \frac{f(u)}u\,du=Ce^{k\sigma t}-\frac{\sigma}kx^k, \qquad C=-C_1\sigma,
\label{eq:112}
\end{equation}
in the construction of which the invariance of equation \eqref{eq:104} with respect to transformation \eqref{eq:105} was taken into account.

\textbf{\textit{Solution 31}.}
In equation \eqref{eq:104}, we set $\zeta=f$ and $\lambda=p(x)f(u)$ (recall that $\lambda$ can be any function dependent on
$x$, $t$, and $u$; see Item 2 in Section~\ref{ss:2.3}). On dividing by $f$, we get
\begin{align}
{-}\vartheta_{t}\frac 1f+(a\vartheta_x)_x +p\vartheta+b\vartheta_x\frac gf+ch-pF=0,
\label{eq:3000}
\end{align}
where $F=\int f(u)\,du$.

Assuming the function $f$ to be given arbitrarily, we look for the functions $g$ and $h$ in the form
\begin{align}
g=f\Bl(k_1+k_2\frac 1f+k_3F\Br), \quad \ h=m_1+m_2\frac 1f+m_3F,
\label{eq:3001}
\end{align}
where $k_i$ and $m_i$
are some constants ($i=1,\,2,\,3$). Substituting \eqref{eq:3001} into \eqref{eq:3000}, we arrive at the equations
\begin{equation}
\begin{aligned}
(a\vartheta_x)_x+p\vartheta+k_1b\vartheta_x+m_1c&=0,\\
-\vartheta_t+k_2b\vartheta_x+m_2c&=0,\\
-p+k_3b\vartheta_x+m_3c&=0,
\end{aligned}
\label{eq:3002}
\end{equation}

Equations \eqref{eq:3002} admit the following exact solution
\begin{align}
k_2=k_3=0,\quad \vartheta=m_2c(x)t+\eta(x),\quad p=m_3c(x),
\label{eq:3003}
\end{align}
where the three functions $a=a(x)$, $b=b(x)$, and $c=c(x)$ are connected by one equation
\begin{align}
(ac'_x)'_x+k_1bc'_x+m_3c^2=0,
\label{eq:3004}
\end{align}
and the function $\eta$ are described by the linear ODE
\begin{align}
(a\eta'_x)'_x+k_1b\eta'_x+m_3c\eta+m_1c=0.
\label{eq:3005}
\end{align}
Note that for given functions $a$ and $c$, equation \eqref{eq:3004} is algebraic with respect to $b$, for given $b$ and $c$
it is a first-order linear ODE with respect to $a$ (which is readily integrated), and for given $a$ and $b$ it is a second-order ODE
with a quadratic nonlinearity with respect to $c$.

To sum up, we have obtained the nonlinear reaction-diffusion type equation
\begin{align}
u_{t}=[a(x)f(u)u_x]_x+b(x)f(u)u_x+c(x)\BL[m_1+\frac{m_2}{f(u)}+m_3\int f(u)\,du\BR],
\label{eq:3006}
\end{align}
where $f(u)$ \arbf,
and any two of the three functions $a=a(x)$, $b=b(x)$, and $c=c(x)$ can be given arbitrarily, while the remaining function satisfies equation \eqref{eq:3004} with $k_1=1$.
Equation \eqref{eq:3006} has the exact solution in implicit form
\begin{align}
\int f(u)\,du=m_2c(x)t+\eta(x),
\label{eq:3007}
\end{align}
where the function $\eta(x)$ is determined by ODE \eqref{eq:3005} with $k_1=1$.

\textit{Remark 7.}
The more general equation
\begin{align*}
u_{t}=[a(x)f(u)u_x]_x+b(x)f(u)u_x+m(x)+\frac{c(x)}{f(u)}+n(x)\int f(u)\,du,
\end{align*}
where $f=f(u)$ and $m=m(x)$ \arbfs,
and the four functions $a=a(x)$, $b=b(x)$, $c=c(x)$, and $n=n(x)$ are connected by one equation
(algebraic in $b$ and $n$, and differential in $a$ and $c$)
\begin{align*}
(ac'_x)'_x+bc'_x+cn=0,
\end{align*}
admits the exact solution
\begin{align*}
\int f(u)\,du=c(x)t+\eta(x),
\end{align*}
with the function $\eta(x)$ determined by the ODE
\begin{align*}
(a\eta'_x)'_x+b\eta'_x+n\eta+m=0.
\end{align*}

\textbf{\textit{Solution 32}.}
Solutions of equation \eqref{eq:3000} can be sought in the form
\begin{align}
g=f\Bl(k_1f^{-1}+k_2\Br),\quad \ h=k_3f^{-1}+k_4,\quad \ F=k_5f^{-1}+k_6,
\label{eq:3006**}
\end{align}
where $k_n$
are some constants; the last relation \eqref{eq:3006**} is used to determine the function $f$.
By setting $k_1=0$, $k_2=1$, $k_5=2$, and $k_6=0$ in \eqref{eq:3006**}, we obtain $f=g=u^{-1/2}$, $h=k_3u^{1/2}+k_4$, and $F=2u^{1/2}$.
The corresponding nonlinear reaction-diffusion type equation
\begin{align}
u_t=[a(x)u^{-1/2}u_x]_x+b(x)u^{-1/2}u_x+c(x)(k_3u^{1/2}+k_4),
\label{eq:3007}
\end{align}
where $a(x)$, $b(x)$, and $c(x)$ \arbfs, while and $k_3$ and $k_4$ \arbs,
has an exact solution in implicit form $F=\xi(x)t+\eta(x)$, which can be expressed in explicit form as
\begin{align}
u=\tfrac14[\xi(x)t+\eta(x)]^2.
\label{eq:3008}
\end{align}
The functions $\xi=\xi(x)$ and $\eta=\eta(x)$ are determined by solving the ordinary differential equations
\begin{equation}
\begin{aligned}
(a\xi^\prime_x)^\prime_x+b\xi^\prime_x+\tfrac12k_3c\xi-\tfrac12 \xi^2&=0,\\
(a\eta^\prime_x)^\prime_x+b\eta^\prime_x+\tfrac12k_3c\eta-\tfrac12 \xi\eta+k_4c&=0.
\end{aligned}
\label{eq:3009**}
\end{equation}

For $c(x)=1$, the first equation \eqref{eq:3009**} can be satisfied if we take $\xi(x)=k_3$.

\textit{Remark 8.}
The equation
\begin{align}
u_t=[a(x)u^{-1/2}u_x]_x+b(x)u^{-1/2}u_x+c(x)u^{1/2}+d(x),
\label{eq:3010}
\end{align}
which is more general than \eqref{eq:3007},
has an exact solution of the form \eqref{eq:3008}.
In the case $d(x)/c(x)=\text{const}$,
equation \eqref{eq:3010} belongs to the class of equations \eqref{eq:07} in question.

\textit{Remark 9.}
The nonlinear delay PDE
\begin{align*}
u_t&=[a_1(x)u^{-1/2}u_x]_x+[a_2(x)w^{-1/2}w_x]_x+b_1(x)u^{-1/2}u_x+b_2(x)w^{-1/2}w_x\\
&+c_1(x)u^{1/2}+c_2(x)w^{1/2}+d(x),\quad \
w=u(x,t-\tau),
\end{align*}
where $\tau$ is the delay time and $a_1(x)$, $a_2(x)$, $b_1(x)$, $b_2(x)$, $c_1(x)$, $c_2(x)$, and $d(x)$ \arbfs,
also admits an exact solution of the form \eqref{eq:3008}.

\textbf{\textit{Solution 33}.}
Now we consider the equation
\begin{align}
-\vartheta_{t}+\vartheta_{xx}f+\vartheta_x^2\Bl(\frac f\zeta \Br)^{\!\prime}_{\!u}-k\Bl(\frac f\zeta \Br)^{\!\prime}_{\!u}\vartheta+k\Bl(\frac f\zeta \Br)^{\!\prime}_{\!u}Z+h\zeta=0.
\label{eq:0888}
\end{align}
which, by virtue of \eqref{eq:02}, is equivalent to equation \eqref{eq:08} for $a=c=1$ and $b=0$.

An exact solution of equation \eqref{eq:0888} is sought in the form
\begin{align*}
\vartheta=Ax^2+Bx+Ce^{-\lambda t},
\end{align*}
where $A$, $B$, $C$, and $\lambda$ are constants to be found.
Omitting the intermediate calculations, we ultimately arrive at the equation
\begin{align}
u_t=[f(u)u_x]_x-\frac12\lambda u-\gamma^2\frac u{f(u)}-\lambda \frac u{f(u)}\int \frac {f(u)}u\,du,
\label{eq:0888}
\end{align}
which has two exact solutions
\begin{align}
\int \frac {f(u)}u\,du=\frac 14\lambda x^2\pm \gamma x+\beta e^{-\lambda t},
\label{eqq:003}
\end{align}
where $\beta$, $\gamma$, and $\lambda$ \arbs.

\textit{Remark 10.}
The described approach also makes it possible to obtain other exact solutions of equation \eqref{eq:07}, which are not considered here
(recall that in this article we only look at nonlinear equations of a fairly general form that depend on arbitrary functions).

\textit{Remark 11.}
This approach can also be used to construct exact solutions of nonlinear ordinary differential equations with variable coefficients.

\subsection{Using transformation (\ref{eq:02}) to simplify equations}\label{ss:3.4}

Transformation \eqref{eq:02} can also be used to simplify nonlinear PDEs.
To illustrate this, we consider the equation
\begin{align}
u_t=au_{xx}+f(u)u^2_x+b(x)g(u)u_x+c(x)h(u),
\label{e500}
\end{align}
where $a$ is a constant.

Transformation \eqref{eq:02} reduces equation \eqref{e500} to the form
\begin{align}
\vartheta_t=a\vartheta_{xx}+\vartheta_x^2\frac 1\zeta\BL[f(u)-a\frac{\zeta'_u}{\zeta}\BR]+b(x)g(u)\vartheta_x+c(x)h(u)\zeta.
\label{e501}
\end{align}
In \eqref{e501}, we set
\begin{align*}
f(u)-a\frac{\zeta'_u}{\zeta}=0,\quad g(u)=1, \quad h(u)\zeta=1.
\end{align*}
Whence
\begin{align*}
\zeta=\exp\BL[\frac 1a\int f(u)\,du\BR],\quad h(u)=\exp\BL[-\frac 1a\int f(u)\,du\BR].
\end{align*}
As a result, we obtain the nonlinear equation
\begin{align}
u_t=au_{xx}+f(u)u^2_x+b(x)u_x+c(x)\exp\BL[-\frac 1a\int f(u)\,du\BR],
\label{e504}
\end{align}
where $b(x)$, $c(x)$, and $f(u)$ \arbfs, which can be reduced with the transformation
\begin{align}
\vartheta=\int \exp\BL[\frac 1a\int f(u)\,du\BR]\,du
\label{e505}
\end{align}
to the linear equation
\begin{align}
\vartheta_t=a\vartheta_{xx}+b(x)\vartheta_x+c(x).
\label{e506}
\end{align}
Some exact solutions of this equation can be found in \cite{polnaz2016}.

\textit{Remark 12.}
Note that in equations \eqref{e504} and \eqref{e506}, the functional coefficients $a(x)$ and $b(x)$ can be replaced with $a(x,t)$ and $b(x,t)$.

\textit{Example 3}.
In the special case $a=1$ and $f(u)=1$, equation \eqref{e504} becomes
\begin{align*}
u_t=u_{xx}+u^2_x+b(x)u_x+c(x)e^{-u},
\end{align*}
and transformation \eqref{e505} can be written in explicit form as $u=\ln\vartheta$.
As a result, we obtain equation \eqref{e506} with $a=1$.

\section{Indirect functional separation of variables}\label{s:4}

\subsection{Functional separation of variables based on the nonclassical method of symmetry reductions}\label{ss:4.1}

The method of functional separation of variables based on
transformation \eqref{eq:02} is closely related to the nonclassical method of symmetry reductions
which is based on an invariant surface condition  \cite{blu1969}.
To show this, we differentiate formula \eqref{eq:02} with respect to $t$ to obtain
\begin{align}
u_t=Q(x,t)\phi(u) \label{eq:3020}
\end{align}
where $Q(x,t)=\vartheta_t$ and $\phi(u)=1/\zeta(u)$.

Relation \eqref{eq:3020} can be treated as a first-order differential constraint or an invariant surface condition of a special form
(in general, an invariant surface condition is a quasilinear first-order PDE of general form),
which can be used to find exact solutions of equation \eqref{eq:07} through a compatibility analysis of
the overdetermined pair of differential equations \eqref{eq:07} and \eqref{eq:3020} with the single unknown~$u$.
The invariant surface condition \eqref{eq:3020} is equivalent to relation \eqref{eq:02};
at the initial stage, both functions $Q(x,t)$ and $\phi(u)$ included on the right-hand side of \eqref{eq:3020} are
considered arbitrary, and the specific form of these functions is determined in the subsequent analysis.

A description of the nonclassical method of symmetry reductions and examples of its application to construct exact solutions of nonlinear PDEs can be found, for example, in
\cite{blu1969,nuc1992,pic1992a,cla1995,olv1996,cla1997,lev1989,puc2000,pol2005,pol2012}.
Although the invariant surface condition \eqref{eq:3020} is equivalent to the functional relation \eqref{eq:02},
the subsequent procedure for finding exact solutions by the nonclassical method of symmetry reductions
(or by the method of differential constraints)
and that by the direct method for seeking functional separable solutions differ significantly.
Let us compare the effectiveness of these methods by the example of the reaction-diffusion type equation \eqref{eq:07}
(since its functional separable solutions have already been obtained in Sections \ref{ss:3.2} and \ref{ss:3.3}).
To construct exact solutions by the nonclassical method of symmetry reductions, we will use relation \eqref{eq:3020} as an invariant surface condition.

\textit{Remark 13}.
The nonclassical method of symmetry reductions, based on the invariant surface condition \eqref{eq:3020},
and the method of differential constraints \cite{yan1964}, based on the single differential constraint \eqref{eq:3020}, end up in the same result.
A description of the method of differential constraints and examples of its application to construct exact solutions of nonlinear PDEs can be found,
for example, in \cite{mel1983,sid1984,gal1994,olv1994,kap1995,kap1998,and1998,kap2003,pol2005,pol2012}.

Taking into account the last remark, below we use the method of differential constraints \cite{pol2012}.
We solve equation \eqref{eq:07} for the highest derivative $u_{xx}$ and eliminate $u_t$ with the help of \eqref{eq:3020} to obtain
\begin{align}
u_{xx}=-\frac {f^\prime_u}fu_x^2-\BL(\frac{a^\prime_x}a+\frac ba\frac gf\BR)u_x+\frac{Q\phi-ch}{af}.
\label{eq:3021}
\end{align}

Differentiating \eqref{eq:3020} twice with respect to $x$ and taking into account relation \eqref{eq:3021}, we get
\begin{equation}
\begin{aligned}
u_{tx}&=Q\phi^\prime_uu_x+Q_x\phi,\\
u_{txx}&=Q\phi'_uu_{xx}+Q\phi''_{uu}u_x^2+2Q_x\phi'_uu_x+Q_{xx}\phi\\
&=Q\BL(\phi''_{u}-\frac{f'_u}{f}\phi'_u\BR)u_x^2+A_1(x,t,u)u_x+A_0(x,t,u),\\
A_1(x,t,u)&=\BL[2Q_x-Q\BL(\frac{a'_x}a+\frac ba\frac gf\BR)\BR]\phi'_u,\\
A_0(x,t,u)&=Q_{xx}\phi-\frac{cQ}a\frac hf\phi'_u+\frac{Q^2}a\frac\phi f\phi'_u.
\end{aligned}
\label{eq:3022}
\end{equation}
where $A_1(x,t,u)$ and $A_0(x,t,u)$ are independent of the derivative $u_x$ and are expressed in terms of the functions appearing in
the original PDE \eqref{eq:07} and the invariant surface condition \eqref{eq:3020}.

Differentiating \eqref{eq:3021} with respect to $t$ and taking into account relation \eqref{eq:3020} and the first formula of \eqref{eq:3022}, we find the
mixed derivative in a different way,
\begin{equation}
\begin{aligned}
{}\!\!\!u_{xxt}&=-Q\BL[\phi\BL(\frac{f'_u}{f}\BR)^{\!\prime}_{\!u}+2\frac{f'_u}{f}\phi'_u\BR]u_x^2+B_1(x,t,u)u_x+B_0(x,t,u),\\
{}\!\!\!B_1(x,t,u)&=-2Q_x\phi\frac{f'_u}{f}-\frac{a_xQ}a\phi'_u-\frac baQ\BL(\frac{g\phi}f\BR)^{\!\prime}_{\!u},\\
{}\!\!\!B_0(x,t,u)&=-\frac{a_xQ_x}a\phi-\frac{bQ_x}a\frac gf\phi-\frac{cQ}a\phi\BL(\frac hf\BR)^{\!\prime}_{\!u}+\frac{Q_t}a\frac\phi
f+\frac{Q^2}a\phi\BL(\frac\phi f\BR)^{\!\prime}_{\!u},
\end{aligned}
\label{eq:3023}
\end{equation}
where $B_1(x,t,u)$ and $B_0(x,t,u)$ are independent of $u_x$.

Equating the third-order mixed derivatives \eqref{eq:3022} and \eqref{eq:3023}, we get the following
relation, quadratic in~$u_x$:
\begin{equation}
Ku_x^2+Mu_x+N=0,
\label{eq:3024}
\end{equation}
where
\begin{align}
\!\!K&=Q\BL[\phi''_{u}+\phi'_u\frac{f'_u}{f}+\phi\BL(\frac{f'_u}{f}\BR)^{\!\prime}_{\!u}\BR],\notag\\
\!\!M&=2Q_x\BL(\phi'_u+\frac{f'_u}{f}\phi\BR)+\frac{bQ}a\BL(\frac{g}{f}\BR)^{\!\prime}_{\!u}\phi,\label{eq:3025}\\
\!\!N&=Q_{xx}\phi+Q_x\phi\BL(\frac{a'_x}a+\frac ba\frac gf\BR)-\frac{Q_t}{a}\frac{\phi}{f}+\frac{cQ}a\BL[\phi\BL(\frac{h}{f}\BR)^{\!\prime}_{\!u}-\frac{h}{f}\phi^\prime_u\BR]
+\frac{Q^2}a\frac{\phi^2f^\prime_u}{f^2}.\notag
\end{align}
The functional coefficients $K$, $M$, and $N$ depend on $a$, $b$, $c$, $f$, $g$, $h$, $Q$, $\phi$ and their derivatives (and are independent of $u_x$).
By equating in \eqref{eq:3024} the functional coefficients $K$, $M$, and $N$ with zero (the procedure of
splitting by the derivative $u_x$), one obtains the determining system of equations $K=0$, $M=0$, $N=0$. Next, we only need the
first equation of this system (corresponding to $K=0$), which, after dividing by $Q$, takes the form
\begin{align}
\phi''_{u}+\phi'_u\frac{f'_u}{f}+\phi\Bl(\frac{f'_u}{f}\Br)^{\!\prime}_{\!u}=0.
\label{eq:3026}
\end{align}
This equation admits the first integral
\begin{align}
\phi'_u+\phi\frac{f'_u}{f}=C_1.
\label{eq:3026a}
\end{align}
Considering $f$ to be an arbitrary function and $\phi$ to be an unknown function and integrating \eqref{eq:3026a}, we find the general solution of
equation \eqref{eq:3026}:
\begin{align}
\phi=\frac 1f\BL(C_1\int f\,du+C_2\BR),
\label{eq:3027}
\end{align}
where $C_1$ and $C_2$ \arbs.
Thus, the nonclassical method of symmetry reductions with the invariant surface condition \eqref{eq:3020} leads to exact solutions in which the functions $f$ and $\phi$
(involved in the original equation and the invariant surface condition) are related by \eqref{eq:3027}.

Using the invariant surface condition \eqref{eq:3020} is equivalent to representing the solution in the form
\eqref{eq:02}. Since $\phi=1/\zeta$, solution \eqref{eq:3027} can be rewritten in terms of $f$ and $\zeta$ as
\begin{align}
\zeta=f\BL(C_1\int f\,du+C_2\BR)^{\!-1}.
\label{eq:3028}
\end{align}

We now consider some solutions obtained in Sections \ref{ss:3.2} and \ref{ss:3.3} by the method of functional separation of variables.
Solution \eqref{eq:801ac} of equation \eqref{eq:801ab} and solution \eqref{eq:112} of equation \eqref{eq:111}
are special cases of solutions \eqref{eq:02} with $\zeta = f /u$. These solutions are different from \eqref{eq:3028};
consequently, they cannot be obtained by the nonclassical method of symmetry reductions with the invariant surface condition \eqref{eq:3020}.
Also, more complex solutions \eqref{eq:14}, \eqref{ee:705}, \eqref{eq:503}, \eqref{eq:508}, \eqref{eq:703}, \eqref{eq:1200}, and \eqref{eq:29}, in which the function $\zeta$ depends not only on $f(u)$
but also on other functional coefficients $g(u)$ or/and $h(u)$
of the considered class of equations \eqref{eq:07}, cannot be obtained by the nonclassical method of symmetry reductions with condition \eqref{eq:3020}.

\textit{Remark 14}. It can be shown that exact solutions listed above cannot be obtained by the nonclassical method of symmetry reductions
using an invariant surface condition of the form $u_t = U(x,t,u)$, which is more general than \eqref{eq:3020}.

\textit{Remark 15}.
Importantly, the vast majority of solutions constructed in this paper are non-invariant
(that is, they cannot be obtained using the classical group analysis of differential equations
\cite{ovs1982,olv1986,ibr1994}).

\subsection{Some remarks on weak symmetries}\label{ss:4.2}

In applying the nonclassical method of symmetry reductions to equation \eqref{eq:07},
the loss of some exact solutions occurred when the splitting procedure in powers of $u_x$ was applied to relation \eqref{eq:3024}--\eqref{eq:3025}.
Theoretically, in order to avoid such losses, we can further search for weak symmetries \cite{puc1992b,dzh1994,vor1996}.
Consider two possible algorithms for finding weak solutions by looking at the example of the nonlinear equation \eqref{eq:07}.

\textbf{\textit{The first algorithm.}} This algorithm consists of two stages.

1.\enspace The first (composite) stage suggests obtaining relation \eqref{eq:3024} and so leads to the same results as applying the nonclassical method
until the splitting procedure in powers of~$u_x$.

2.\enspace The second stage suggests analysing three PDEs \eqref{eq:3020}, \eqref{eq:3021}, and \eqref{eq:3024}--\eqref{eq:3025} for consistency (in order to obtain the
determining equation, which must then be integrated).

The compatibility analysis of these PDEs is carried out as follows. Equation \eqref{eq:3024} is differentiated with respect to $t$, after which the derivatives
$u_t$ and $u_{xt}$ are eliminated from the resulting expression using relation \eqref{eq:3020} and the first formula of \eqref{eq:3022}.
As a result, we obtain
\begin{equation}
Pu_x^2+Qu_x+R=0,
\label{eq:2200}
\end{equation}
where
\begin{equation}
\begin{aligned}
P&=K_t+UK_u+2U_uK,\\
Q&=M_t+UM_u+U_uM+2U_xK,\\
R&=N_t+UN_u+U_xM.
\end{aligned}
\label{eq:2201}
\end{equation}
For brevity, short notations are used here:
$$
K=K(x,t,u),\quad M=M(x,t,u),\quad N=N(x,t,u),\quad U=Q(x,t)\phi(u).
$$
Having further eliminated the derivative $u_x$ from equations \eqref{eq:3024} and \eqref{eq:2200}, we obtain the determining equation, which in the
nondegenerate case ($MP-KQ\not\equiv 0$) has the form
\begin{equation}
K(NP-KR)^2-M(MP-KQ)(NP-KR)+N(MP-KQ)^2=0.
\label{eq:2202}
\end{equation}

Equation \eqref{eq:2202} is a very complex nonlinear PDE, which includes third-order derivatives ${Q}_{xxt}$ and $\phi'''_{uuu}$ (recall that ${Q}$ and $\phi$
are both   unknown functions), whose length in expanded form (taking into account the relations \eqref{eq:3024} and \eqref{eq:2201}) occupies almost an entire
page. In addition, equation \eqref{eq:2202}, which includes one or more arbitrary functions $f(u)$, $g(u)$, etc.,
must be solved together with the equations \eqref{eq:3020} and \eqref{eq:3021} (or the
original equation). As a result, instead of one equation \eqref{eq:07}
(or equation \eqref{eq:08} together with \eqref{eq:02}),
it is necessary in this case to deal with a much more complex system of coupled nonlinear PDEs.

\textit{Example 4}.
For greater clarity, let us look at the linear heat equation $u_t=u_{xx}$, which is obtained from \eqref{eq:07} by setting
$$
a(x)=1,\quad b(x)=0,\quad c(x)=0,\quad f(u)=1.
$$
In this case, one has to substitute into equation \eqref{eq:2202} the following functions:
\begin{equation}
\begin{aligned}
K&=U_{uu},\quad M=2U_{xu},\quad N=U_{xx}-U_t,\quad U={Q}\phi;\\
P&=U_{tuu}+UU_{uuu}+2U_uU_{uu},\\
Q&=2(U_{xtu}+UU_{xuu}+U_uU_{xu}+U_xU_{uu},\\
R&=U_{xxt}-U_{tt}+U(U_{xxu}-U_{tu})+2U_xU_{xu}.
\end{aligned}
\label{eq:2202*}
\end{equation}
One can see that the nonlinear third-order equation \eqref{eq:2202}--\eqref{eq:2202*} becomes isolated (can be solved independently of the original equation);
it is far more complicated than the linear heat equation under consideration.

The degenerate case of $MP-KQ\equiv 0$ can be treated likewise.
\medskip

It is apparent from the above examples that the method in question, based on the analysis of three PDEs \eqref{eq:3020}, \eqref{eq:3021}, and \eqref{eq:3024}, is
extremely difficult for practical use.

\textbf{\textit{The second algorithm.}}
In this case, we differentiate formula \eqref{eq:02} with respect to $t$ and $x$.
As a result, we obtain two relations
\begin{align}
u_t=\vartheta_t\phi(u),\quad \  u_x=\vartheta_x\phi(u),
\label{eq:7010}
\end{align}
which can be interpreted as two compatible differential constraints,
where the functions $\vartheta=\vartheta(x,t)$ and $\phi(u)=1/\zeta(u)$ are to be determined.
Differentiating the second relation \eqref{eq:7010} with respect to $x$, we find the second derivative
\begin{align}
u_{xx}=\vartheta_{xx}\phi+\vartheta_x\phi^\prime_uu_x=\vartheta_{xx}\phi+\vartheta_x^2\phi\phi^\prime_u,\quad \ \phi=\phi(u).
\label{eq:7011}
\end{align}

Next, we substitute the derivatives \eqref{eq:7010} and \eqref{eq:7011} in \eqref{eq:07}.
As a result, we arrive at an equation that is equivalent to equation \eqref{eq:08}.
Using further the generalized splitting principle described in Section~\ref{ss:2.1}, one can find the exact solutions obtained in Section~\ref{ss:3.2}.
However, it will not be possible to find the solutions obtained in Section \ref{ss:3.3} in this way.
In order to find these solutions, one must first integrate the differential relations \eqref{eq:7010} and return to the original relation \eqref{eq:02},
and then consider the equivalent equations described in Section~\ref{ss:2.3}.

Thus, it seems that the use of transformation \eqref{eq:02} is more effective for constructing exact solutions
than the use of one or two equivalent differential constraints.

\section{Brief conclusions}\label{s:5}

A general method for constructing exact solutions of nonlinear PDEs has been described,
which is based on nonlinear transformations with an integral term in combination with the generalized splitting principle.
The high productivity of the method has been illustrated by nonlinear equations of the reaction-diffusion type with variable
coefficients that depend on one, two or three arbitrary functions.
Many new exact functional separable solutions and
generalized traveling wave solutions have been obtained.
The effectiveness of various methods for constructing exact solutions of nonlinear differential equations has been compared.

The direct method of functional separation of variables based on transformation \eqref{eq:02}, in addition to diffusion-type equations,
is also applicable to other classes of PDEs.
In particular, these include nonlinear wave equations, nonlinear Klein--Gordon type equations, nonlinear telegraph-type equations, and others;
these also include some third- and higher-order
PDEs. This method is easy to generalize to equations with three or more independent variables.

\section*{Acknowledgments}

The study was supported by the Ministry of Eduction and Science of the Russian Federation within the framework of the State Assignment (Registration
Number AAAA-A17-117021310385-6) and partially supported by the Russian Foundation for Basic Research (project No.\ 18-29-03228).

I wish to express my deep gratitude to Alexei Zhurov for fruitful discussions that helped improve this paper.

\renewcommand{\refname}{References}

\end{document}